\documentclass[a4paper,12pt]{article}
\usepackage{amsmath,amssymb}
\usepackage{authblk}
\usepackage{graphicx}
\usepackage{subcaption}
\usepackage{blindtext}
\usepackage[title]{appendix}
\allowdisplaybreaks[4]

\begin{document}
\title{Nucleon matrix elements of axial anomaly, axial currents and pseudoscalar currents in the QCD sum rule}

\author{Janardan Prasad Singh\footnote{janardanmsu@yahoo.com}  \\ {\scriptsize Physics Department, Faculty of Science, The Maharaja Sayajirao University of Baroda, Vadodara-390002, Gujarat, India}}

\maketitle
\abstract{ We have analyzed one-nucleon matrix elements of current-current correlators; the currents consist of pseudoscalar octet, isovector and isoscalar currents, axial anomaly, and axial isovector and isoscalar currents. Using QCD sum rules, the coupling constants of nucleon with each of these currents have been expressed in terms of nucleon matrix elements of quark, gluon and quark-gluon composite operators and moments of parton distribution function. On the phenomenological side, contribution from the non-diagonal matrix elements of operators between nucleon and its excited states or continuum states have also been  accounted for. For the pseudoscalar coupling constants of the nucleon two expressions have been obtained in which one of them consists of only moments of parton distribution function but yielding approximately same numerical result as the other one. Of particular interest is the nucleon matrix element of axial anomaly which has been largely ignored in the current literature. }

\section{Introduction}
The axial currents have an essential role in understanding nucleon and hyperon decays. The nucleon matrix elements of axial currents at zero momentum transfer are called  axial coupling constants. Since the axial current $\bar{q}(x)\gamma_{\mu}\gamma_5q(x)$ is a spin operator and its nucleon matrix element gives axial charge 
\begin{equation}
2Ms_{\mu}\Delta q= \langle p,s| \bar{q}(x)\gamma_{\mu}\gamma_5q(x)|p,s \rangle
\end{equation}
the parton model interprets $\Delta q $ (without taking into account gluonic effects in  the flavour-singlet channel) as the fraction of proton's spin carried by the quark and antiquark of flavour q. The isovector, isoscalar and octet axial charges are : $g_A^{u-d}=\Delta u-\Delta d, g_A^{u+d}=\Delta u+\Delta d, g_A^8=\Delta u+\Delta d-2\Delta s$. The singlet axial charges have anomalous dimension and hence $g_A^{u+d}$ has renormalization scale dependence. $g_A^{u-d}$ and $g_A^{u+d}$ are required in Bjorken sum rule \cite{Bjorken69}  and Ellis-Jaffe sum rule \cite{Ellis} respectively. Goldberger-Trieman relation and its generalization relate pseudoscalar couplings of a nucleon with mesons and their decay constants to the corresponding axial coupling constants of the nucleon. In case of singlet axial current, the   Goldberger-Trieman relation also involves the proper vertex for the coupling of the gluonic topological charge density to the nucleon \cite{Narison, Shore}. \\
\par \hspace{12pt}Unlike the case of quark axial current, there is no gauge invariant local operator for gluons which gives helicity contribution $\Delta G$ to the nucleon spin. For the singlet axial current $ J^0_{\mu 5}=\sum_f \bar{\psi_f}\gamma_{\mu}\gamma_5\psi_f $, we have well known divergence relation
\begin{equation}  \partial^{\mu}J^{0}_{\mu5}=\sum_q 2m_q\bar{\Psi}_qi\gamma_5\Psi+\frac{n_f\alpha_s}{4\pi}G^a_{\mu \nu}\tilde{G}^{a\mu \nu}  \end{equation}
where axial anomaly appears. Here $\tilde{G}^{a\mu \nu} =\frac{1}{2}\epsilon^{\mu\nu\rho \sigma}G^a_{\mu \nu}$ and $\epsilon^{0123}=-1$. To our knowledge, the nucleon matrix element of the axial anomaly $Q=\frac{\alpha_s}{\pi}G^a_{\mu \nu}\tilde{G}^{a\mu \nu}$, also known as topological charge density, has not been investigated except in the chiral limit \cite{Kuhn1, Weiss}. Perhaps this is mainly because its direct connection to $\Delta G$ has not been demonstrated. However, $\Delta G$ can be  related to  operators which can be reduced, by taking local limit or by taking divergence, to axial anomaly. Nucleon matrix element of (i) $ G^{+\alpha}_a (z^-)L^{ab}(z^-,0)\tilde{G}^+_{ \alpha, b}(0) $ where $ L^{ab}(z^-,0) $ is a light-cone gauge link  \cite{Manohar} and (ii) $K^+$,  where $K^{\mu}=\epsilon^{\mu\nu\rho\lambda}(A^a_{\nu}F^a_{\rho\lambda}+\frac{g_s}{3}f^{abc}A^a_{\nu}A^b_{\rho}A^c_{\lambda}) $ is topological current whose divergence is axial anomaly \cite{Manohar, Jaffe}, can be related to $\Delta G$. In light-cone gauge both of these operators become gluon spin operator\cite{Jaffe}. In our approach of the theoretical investigation of isosinglet axial coupling $g_A^{u+d}$ and strange axial coupling $g_A^s$,  knowledge of nucleon matrix element of axial anomaly becomes indispensable. It has been pointed out in \cite{Gross} that if the axcial anomaly is neglected then one expects a large violation of isospin symmetry by the order of $(m_u-m_d)/(m_u+m_d)$ which is around 30\% in Bjorken electroproduction sum rule\cite{Bjorken69}. In last close  to four decades there have been several experiments including those from which neutron data is also avilable where Bjorken sum rule has been tested. From Ref. \cite{Adolph} we learn 
\begin{equation} \hat{\Gamma}^p- \hat{\Gamma}^n=0.18\pm 0.008\pm 0.014  \end{equation}
which agrees with the Bjorken prediction and there is no sign of large violation of isospin in data. In Eq. (3) $\hat{\Gamma}^{p,n} $ is the first moment of the respective first spin-dependent structure function $g_1^{p,n} $. Next, consider the Ellis-Jaffe sum rule \cite{Ellis} which assumes that the strange quark "sea" in a nucleon is unpolarized implying \\
\begin{equation}
 \langle p,s| \bar{s}(x)\gamma_{\mu}\gamma_5s(x)|p,s \rangle=0
\end{equation}
which is apparantly justified by OZI rule. This means that the matrix element of octet axial current can be written  as
\begin{equation}
\begin{aligned}
 \langle p,s| \bar{u}(x)\gamma_{\mu}\gamma_5u(x)+\bar{d}(x)\gamma_{\mu}\gamma_5d(x)-2 \bar{s}(x)\gamma_{\mu}\gamma_5s(x)|p,s \rangle \\
=  \langle p,s| \bar{u}(x)\gamma_{\mu}\gamma_5u(x)+\bar{d}(x)\gamma_{\mu}\gamma_5d(x)|p,s \rangle
\end{aligned}
\end{equation}
But isosinglet axial current $\bar{u}\gamma_{\mu}\gamma_5u+\bar{d}\gamma_{\mu}\gamma_5d $ has an anomaly. If anomaly is neglected, as was noted above, one expects large violation of isospin and flavor $SU(3)$ symmetry. Thus the Ellis-Jaffe sum rule \cite{Ellis} is bound to fail as confirmed by the experiment \cite{Adolph}. Inclusion of anomaly is important to underastand the experimental data.\\ 
\par \hspace{12pt}The standard QCD sum rule for calculating axial coupling constants uses external fields. Two-point correlation function of nucleon interpolating fields is evaluated in the presence of weak axial vector field \cite{Belyaev1, Belyaev85, Chiu, Gupta, Pasupathy}. The limits on the part of proton spin carried by light quarks and the derivative of the QCD susceptibility have been found from self-consistency of the sum rule \cite{Ioffe98}. Belitsky and Teryaev  \cite{Teryaev} considered a three-point function of nucleon interpolating   fields and the divergence of singlet axial current . The form factor $g_A^0(q^2)$ is related to vacuum condensates of quarl-gluon composite operators through a double dispersion relation. In this approach, the extrapolation  to  $g_A^0(0)$ involve large uncertainties. In the third approach by Nishikawa et al. \cite{Nishikawa, Nishikawa1, JPS1}, the two-point correlation function of axial currents in one-nucleon state is evaluated. Here, the axial coupling constants of the nucleon are expressed in terms of $\pi-N$ and $K-N$ sigma terms, and moment of parton distributions. The  perturbative contribution is subtracted from the beginning and the continuum can be reduced to a small value \cite{Nishikawa, Nishikawa1}. In the present work, for calculating nucleon matrix elements of axial, pseudoscalar currents and axial anomaly, we follow this third approach. \\
\par \hspace{12pt}There has been attempts to determine nucleon matrix element of  axial anomaly using lattice QCD\cite{Mandula, Mandula1}.  However, no reliable signal was found on using existing set of quenched lattices\cite{Mandula1}. .
\par \hspace{12pt} In sec. 2, we calculate nucleon matrix elements of octet pseudoscalar current and axial anomaly from nucleon matrix element of pseudoscalar-pseudoscalar current correlator and axial anomaly-pseudoscalar current correlator. These correlators have been, in turn, evaluated in terms of nucleon matrix elements of suitable axial-axial current correlators. In sec. 3, we calculate nucleon matrix elements of isovector and isoscalar axial currents, and s-quark axial current from nucleon matrix elements of the suitable axial-axial current correlators. In sec. 4, we calculate nucleon matrix elements of isovector and isoscalar pseudoscalar currents from nucleon matrix elements of the corresponding pseudoscalar-pseudoscalar current correlators which, in turn, have been derived from  the nucleon matrix elements of the suitable axial-axial current correlators. In sec.5, we present a summary of the results obtained, check their  consistency and compare them with those in the literature. In the Appendix, we give some details of a couple of steps used in the text.

\section{Octet pseudoscalar current and axial anomaly}
 Let us define SU(3) octet and singlet axial currents and octet pseudoscalar currents as : 
\begin{equation}
\begin{aligned}
J^{(8)}_{\mu5}=\frac{1}{\sqrt{6}}( \bar{u}\gamma_{\mu}\gamma_5 u+\bar{d}\gamma_{\mu}\gamma_5 d-2\bar{s}\gamma_{\mu}\gamma_5s)\\
J^{(0)}_{\mu5}=\frac{1}{\sqrt{3}}( \bar{u}\gamma_{\mu}\gamma_5 u+\bar{d}\gamma_{\mu}\gamma_5 d+\bar{s}\gamma_{\mu}\gamma_5s)\\
P_8=m_u\bar{u}i\gamma_5 u+m_d\bar{d}i\gamma_5 d-2m_s\bar{s}i\gamma_5s
\end{aligned}
\end{equation}
\begin{equation} Q=\frac{\alpha_s}{\pi}G\tilde{G}, \partial^{\mu}J^{(8)}_{\mu5}=\frac{2}{\sqrt{6}}P_8, \partial^{\mu}J^{(0)}_{\mu5}=\frac{1}{\sqrt{3}}\big[\sum_q 2m_q\bar{\Psi}_qi\gamma_5\Psi+\frac{3}{4}Q\big]  \end{equation}
Consider the following correlation functions:
\begin{equation}
\begin{aligned}
\Pi^{(88)}(q^2) \equiv i \int d^4xe^{iqx}\langle T\{P_8(x),P_8(0)\}\rangle_N=\frac{3}{2}q^{\mu}q^{\nu} i \int d^4xe^{iqx}\langle T\{J^{(8)}_{\mu5}(x), J^{(8)}_{\nu5}(0) \}\rangle_N \\
-\langle N(P)|m_u\bar{u}(0)u(0)+m_d\bar{d}(0)d(0)+4m_s\bar{s}(0)s(0)|N(P)\rangle
\end{aligned}
\end{equation}

\begin{equation}
\begin{aligned}
\Pi^{(08)}(q^2) \equiv  i \int d^4xe^{iqx}\langle T\{Q(x),P_8(0)\}\rangle_N \cong q^{\mu}q^{\nu} i \int d^4xe^{iqx}\langle T\{2\sqrt{2}J^{(0)}_{\mu5}(x), J^{(8)}_{\nu5}(0)\} \\-T\{J^{(8)}_{\mu5}(x), J^{(8)}_{\nu5}(0)  \}\rangle_N +8m_s^2 i \int d^4xe^{iqx}\langle T\{\bar{s}(x)i\gamma_5s(x),\bar{s}(0)i\gamma_5s(0)\}\rangle_N \\
-2\langle N(P)|m_u\bar{u}(0)u(0)+m_d\bar{d}(0)d(0)-4m_s\bar{s}(0)s(0)|N(P)\rangle ,
\end{aligned}
\end{equation}
where
\begin{equation}\langle ... \rangle_N \equiv \frac{1}{2} \sum_S [\langle PS|...|PS\rangle-\langle ... \rangle_0 \langle PS|PS\rangle]  \end{equation}
We define the form factors for $P_8$ and $Q$ as
\begin{equation}
\begin{aligned}
\langle P,S|P_8(0)|P+q,S' \rangle =2M \chi_8(q^2)\bar{u}(P,S)i\gamma_5 u(P+q,S')\\
\langle P,S|Q(0)|P+q,S' \rangle =2M \chi_g(q^2)\bar{u}(P,S)i\gamma_5 u(P+q,S')
\end{aligned}
\end{equation}
 We use the normalization $\bar{u}u=1$ and hence, $\chi_8$  and  $\chi_g$ are dimensionless. We will use a reference frame in which $P^{\mu}=(M,\vec{0})$ and call $q^{\mu}=(\omega,\vec{q})$.

\par \hspace{12pt} We can express the spectral function for $\Pi^{(88)}(q^2)$ as :
\begin{equation}
\begin{aligned}
\rho_{88}(\omega, \vec{q})=\frac{-1}{\pi} Im \Pi^{88}(\omega, \vec{q})  \\
=\frac{-2M^2\omega}{M+\frac{\vec{q}^2}{2M}} [\chi_8(-\vec{q}^2)]^2\Big[\delta \big(\omega+\frac{\vec{q}^2}{2M}\big)+\delta\big(\omega-\frac{\vec{q}^2}{2M}\big) \Big] -\\
\sum_i \frac{M\vec{q}^2}{M_i+\frac{\vec{q}^2}{2M_i}} [\chi_{8i}(q^2_{i\pm})]^2\Big[\delta\big(\omega+M-M_i-\frac{\vec{q}^2}{2M_i}\big)-\delta\big(\omega-M+M_i+\frac{\vec{q}^2}{2M_i}\big) \Big]
\end{aligned}
\end{equation}
\begin{equation}q^2=\omega^2-\vec{q}^2, q_{i \pm}^2=(M_i-M)^2-\frac{M}{M_i} \vec{q}^2+O(\vec{q}^4) \end{equation}
$\chi_{8i}$ arises from non-diagonal matrix element between one-nucleon state and and an excited state of a nucleon or continuum states . In terms of $\rho_{88}(\omega, \vec{q})$, we can write the phenomenological expression for $\Pi^{(88)}_{ph}(\omega, \vec{q})$ using Lehmann representation as
\begin{equation}\Pi^{(88)}_{ph}(\omega, \vec{q})=\int_{-\infty}^{+\infty}d\omega'\frac{\rho_{88}(\omega',\vec{q})}{\omega-\omega'}\end{equation}
\vspace{2ex}
Borel transform will be done with respect to $\omega^2$ : \\
 \vspace{-1ex}
  B  $\equiv$  lim  $\frac{(-\omega^2)^{n+1}}{n!}\Big[-\frac{d}{d(-\omega^2)}\Big]^n$ \\
\vspace{-1ex}
\hspace{18pt} $-\omega^2\rightarrow \infty$    \\
\vspace{-1ex}
\hspace{20pt} $n \rightarrow \infty$,  $ -\omega^2/n =s$   \\

For the even part of $\Pi^{(88)}_{ph}(\omega, \vec{q})$ we get

\begin{equation}
\begin{aligned}
B\big[\Pi^{(88)}_{ph}(\omega, \vec{q})\big]=-2\int_{0}^{\infty}d\omega' \omega' \rho_{88}(\omega',\vec{q})e^{-\omega'^2/s}\\
=2M\vec{q}^2\sum_i \frac{M_i-M}{M_i}(\chi_{8i})^2e^{-(M_i-M)^2/s}\\
+\frac{\vec{q}^4}{M}(\chi_8)^2+ 2M\vec{q}^4\sum_i(\chi_{8i})^2e^{-(M_i-M)^2/s}\Big[\frac{M}{2M_i^3}-\\
\frac{2M(M_i-M)}{M_i^2}\frac{\chi'_{8i}}{\chi_{8i}} -\frac{(M_i-M)^2}{sM_i^2}\Big] + terms \, O(\vec{q}^6)
\end{aligned}
\end{equation}

In a similar way, we express the spectral function for $\Pi^{(08)}(q^2)$ as :
\begin{equation}
\begin{aligned}
\rho_{08}(\omega, \vec{q})=\frac{-1}{\pi} Im \Pi^{08}(\omega, \vec{q})  \\
=\frac{-2M^2\omega}{M+\frac{\vec{q}^2}{2M}} [\chi_g(-\vec{q}^2)\chi_8(-\vec{q}^2)]\big[ \delta\big(\omega+\frac{\vec{q}^2}{2M}\big)+\delta\big(\omega-\frac{\vec{q}^2}{2M}\big) \big] -\\
\sum_i \frac{M\vec{q}^2}{M_i+\frac{\vec{q}^2}{2M_i}} [\chi_{gi}(q^2_{i\pm})\chi_{8i}(q^2_{i\pm})]\big[ \delta\big(\omega+M-M_i-\frac{\vec{q}^2}{2M_i}\big)-\delta\big(\omega-M+M_i+\frac{\vec{q}^2}{2M_i}\big) \big]
\end{aligned}
\end{equation}
\begin{equation}\Pi^{(08)}_{ph}(\omega, \vec{q})=\int_{-\infty}^{+\infty}d\omega'\frac{\rho_{08}(\omega',\vec{q})}{\omega-\omega'}
\end{equation}

\begin{equation}
\begin{aligned}
B\big[\Pi^{(08)}_{ph}(\omega, \vec{q})\big]=-2\int_{0}^{\infty}d\omega' \omega' \rho_{08}(\omega',\vec{q})e^{-\omega'^2/s}\\
=2M\vec{q}^2\sum_i \frac{M_i-M}{M_i}\chi_{gi}\chi_{8i}e^{-(M_i-M)^2/s}\\
+\frac{\vec{q}^4}{M}\chi_g\chi_8+ 2M\vec{q}^4\sum_i\chi_{gi}\chi_{8i}e^{-(M_i-M)^2/s}\Big[\frac{M}{2M_i^3}-\\
\frac{M(M_i-M)}{M_i^2}\big(\frac{\chi'_{gi}}{\chi_{gi}}+\frac{\chi'_{8i}}{\chi_{8i}}\big) -\frac{(M_i-M)^2}{sM_i^2}\Big] + terms \, O(\vec{q}^6)
\end{aligned}
\end{equation}

Next, we write below the expressions that come from operator product expansion (OPE) of $\Pi^{(88)}$ and $\Pi^{(08)}$:
\begin{equation}
\begin{aligned}
\Pi^{(88)}_{OPE}(q,P) \equiv  i \int d^4xe^{iqx}\ \langle T\{P_8(x),P_8(0)\}\rangle_N\\
=\frac{1}{2}(m_u \langle\bar{u}u\rangle_N + m_d\langle\bar{d}d\rangle_N +4m_s \langle\bar{s}s\rangle_N) \\
+\frac{\pi \alpha_s}{6q^2}\big[\langle S(\bar{u}\gamma_{\mu}\lambda^au+\bar{d}\gamma_{\mu}\lambda^ad+4\bar{s}\gamma_{\mu}\lambda^as)\sum_q\bar{\psi}_q\gamma_{\nu}\lambda^a\psi_q\rangle_N \big]\big(\frac{q^{\mu}q^{\nu}}{q^2}-g^{\mu \nu}\big)\\
-4i\frac{q^{\mu}q^{\nu}q^{\lambda}q^{\sigma}}{q^6}\langle(\bar{u}S\gamma_{\sigma}D_{\mu}D_{\nu}D_{\lambda}u+\bar{d}S\gamma_{\sigma}D_{\mu}D_{\nu}D_{\lambda}d+4\bar{s}S\gamma_{\sigma}D_{\mu}D_{\nu}D_{\lambda}s)\rangle_N\\
+\frac{1}{q^4}\Big[\frac{1}{36}\langle(\alpha_sG^2)^2\rangle_N \big(\frac{7}{6}+\gamma_E+ln\frac{-q^2}{4}\big) -\frac{4\pi}{27}m_s\langle\bar{s}s\alpha_sG^2\rangle_N \\
-32\pi \alpha_sm_s^2\langle (\bar{s}t^a\gamma_5s)(\bar{s}t^a\gamma_5s)\rangle_N\Big]\\
+16i\frac{q^{\mu}q^{\nu}q^{\lambda}q^{\sigma}q^{\rho}q^{\alpha}}{q^{10}}\Big[\langle \bar{u}S\gamma_{\alpha}D_{\mu}D_{\nu}D_{\lambda}D_{\sigma}D_{\rho}u+\bar{d}S\gamma_{\alpha}D_{\mu}D_{\nu}D_{\lambda}D_{\sigma}D_{\rho}d\\ +4\bar{s}S\gamma_{\alpha}D_{\mu}D_{\nu}D_{\lambda}D_{\sigma}D_{\rho}s\rangle_N\Big]
\end{aligned}
\end{equation}

\begin{equation}
\begin{aligned}
\Pi^{(08)}_{OPE}(q,P) \equiv  i \int d^4xe^{iqx}\ \langle T\{Q(x),P_8(0)\}\rangle_N\\
=m_u \langle\bar{u}u\rangle_N + m_d\langle\bar{d}d\rangle_N -4m_s \langle\bar{s}s\rangle_N \\
+\frac{\pi \alpha_s}{3q^2}\big[\langle S(\bar{u}\gamma_{\mu}\lambda^au+\bar{d}\gamma_{\mu}\lambda^ad-4\bar{s}\gamma_{\mu}\lambda^as)\sum_q\bar{\psi}_q\gamma_{\nu}\lambda^a\psi_q\rangle_N \big]\Big(\frac{q^{\mu}q^{\nu}}{q^2}-g^{\mu \nu}\Big)\\
-8i\frac{q^{\mu}q^{\nu}q^{\lambda}q^{\sigma}}{q^6}\langle(\bar{u}S\gamma_{\sigma}D_{\mu}D_{\nu}D_{\lambda}u+\bar{d}S\gamma_{\sigma}D_{\mu}D_{\nu}D_{\lambda}d-4\bar{s}S\gamma_{\sigma}D_{\mu}D_{\nu}D_{\lambda}s)\rangle_N\\
+8m_s^2\Big[-\frac{2m_s}{q^2}\langle\bar{s}s\rangle_N+4i\frac{q^{\mu}q^{\nu}}{q^4}\langle\bar{s}S\gamma_{\mu}D_{\nu}s\rangle_N-\frac{1}{8q^2}\langle\frac{\alpha_s}{\pi}G^2\rangle_N-\frac{2\pi \alpha_s}{q^4}\langle(\bar{s}\lambda^a\sigma^{\mu \nu}s)( \bar{s}\lambda^a\sigma_{\mu \nu}s)\rangle_N\\ +4\pi\alpha_s\frac{q_{\mu}q^{\nu}}{q^6}\langle(\bar{s}\lambda^a\sigma^{\alpha \mu}s)( \bar{s}\lambda^a\sigma_{\alpha \nu}s)\rangle_N-\frac{4\pi \alpha_s}{3q^4}\langle S(\bar{s}\gamma_{\mu}\lambda^as\sum_q\bar{\psi}_q\gamma_{\nu}\lambda^a\psi_q)\rangle_N\Big(\frac{q^{\mu}q^{\nu}}{q^2}-\frac{3}{4}g^{\mu \nu}\Big)\\
-16i\frac{q^{\mu}q^{\nu}q^{\lambda}q^{\sigma}}{q^8}\langle\bar{s}S\gamma_{\sigma}D_{\mu}D_{\nu}D_{\lambda}s \rangle_N\Big]\\
+\frac{1}{q^4}\Big[-\frac{1}{54}\langle(\alpha_sG^2)^2\rangle_N \big(\frac{7}{6}+\gamma_E+ln\frac{-q^2}{4}\big) +\frac{8\pi}{27}m_s\langle\bar{s}s\alpha_sG^2\rangle_N \\
+64\pi \alpha_sm_s^2\langle (\bar{s}t^a\gamma_5s)(\bar{s}t^a\gamma_5s)\rangle_N\Big]\\
+32i\frac{q^{\mu}q^{\nu}q^{\lambda}q^{\sigma}q^{\rho}q^{\alpha}}{q^{10}}\langle \big[\bar{u}S\gamma_{\alpha}D_{\mu}D_{\nu}D_{\lambda}D_{\sigma}D_{\rho}u+\bar{d}S\gamma_{\alpha}D_{\mu}D_{\nu}D_{\lambda}D_{\sigma}D_{\rho}d\\ -4\bar{s}S\gamma_{\alpha}D_{\mu}D_{\nu}D_{\lambda}D_{\sigma}D_{\rho}s\big]\rangle_N
\end{aligned}
\end{equation}
Nucleon matrix elements which contain covariant derivatives can be expressed as  $\langle \bar{q}S\gamma_{\mu1}D_{\mu2}....D_{\mu n}q\rangle_N=(-i)^{n-1}A_n^q S[P_{\mu1}P_{\mu2}....P_{\mu n}]$ where $A_n^q$ is the nth moment of the parton distribution function \cite{Hatsuda}. Here as well as in Eqs. (19) and (20) S makes the operators symmetric and traceless with respect to the Lorentz indices. This results in
\begin{equation}
\begin{aligned}
i\frac{q^{\mu}q^{\nu}}{q^4}\langle \bar{u}S\gamma_{\mu}D_{\nu}u\rangle_N=\frac{M^2}{2M}A^u_2\big(\frac{\omega^2}{q^4}-\frac{1}{4q^2}\big),\\
-i\frac{q^{\mu}q^{\nu}q^{\lambda}q^{\sigma}}{q^6}\langle \bar{u}S\gamma_{\sigma}D_{\mu}D_{\nu}D_{\lambda}u\rangle_N=\frac{M^4}{2M}A^u_4\big(\frac{\omega^4}{q^6}-\frac{3}{4}\frac{\omega^2}{q^4}+\frac{1}{16q^2}\big),\\
i\frac{q^{\mu}q^{\nu}q^{\lambda}q^{\sigma}q^{\rho}q^{\alpha}}{q^{10}}\langle \bar{u}S\gamma_{\alpha}D_{\mu}D_{\nu}D_{\lambda}D_{\sigma}D_{\rho}u\rangle_N \\
=\frac{M^6}{2M}A^u_6\big(\frac{\omega^6}{q^{10}}-\frac{5}{4}\frac{\omega^4}{q^8}+\frac{15}{32}\frac{\omega^2}{q^6}-\frac{5}{256 q^4}\big).
\end{aligned}
\end{equation}

Next, we take Borel transform of $\Pi^{(88)}_{OPE}(\omega, \vec{q})$ :
\begin{equation}
\begin{aligned}
B\big[\Pi^{(88)}_{OPE}(\omega, \vec{q})\big]=-\frac{4}{9}\pi \alpha_s e^{-\vec{q}^2/s}[\langle \bar{u}u \rangle_0(\langle \bar{u}u \rangle_N+ \langle \bar{d}d \rangle_N)+4\langle \bar{s}s \rangle_0\langle \bar{s}s \rangle_N]\\
+2M^3\Big(-\frac{5}{16}+\frac{25}{16}\frac{\vec{q}^2}{s}-\frac{61}{32}\frac{\vec{q}^4}{s^2}+ terms \, O(\vec{q}^6)\Big)(A^u_4+A^d_4+4A^s_4)\\
+\frac{\pi^2}{6s}e^{-\vec{q}^2/s}\langle(\frac{\alpha_s}{\pi}G^2)^2\rangle_N \frac{1}{6}\big(\frac{13}{6}+ln\frac{s}{4}\big)-\frac{1}{s}e^{-\vec{q}^2/s}\Big[\frac{4\pi}{27}\langle m_s\bar{s}s\alpha_sG^2\rangle_N \\ +32\pi\alpha_sm_s^2\langle (\bar{s}t^a\gamma_5s)(\bar{s}t^a\gamma_5s) \rangle_N \Big]\\ + \frac{M^5}{32}\Big(\frac{51}{s}-175\frac{\vec{q}^2}{s^2}+\frac{1345}{6}\frac{\vec{q}^4}{s^3}+ terms \, O(\vec{q}^6)\Big)(A^u_6+A^d_6+4A^s_6)
\end{aligned}
\end{equation}
where we have also used factorization approximation
\begin{equation}
\langle \bar{q}_f \lambda^a\gamma_{\mu}q_f  \bar{q}_{f'} \lambda^a\gamma_{\nu}q_{f'}\rangle_N \approx-\frac{8}{9}g_{\mu\nu}\langle \bar{q}_f q_f\rangle_0\langle \bar{q}_f q_f\rangle_N \delta_{ff'}.
\end{equation}

By equating coefficients of different powers of  $\vec{q}^2$ from  $B\big[\Pi^{(88)}_{ph}\big]$ and $B\big[\Pi^{(88)}_{OPE}\big]$, we can get different equations. From equating coefficients of $(\vec{q}^2)^0$ we get :
\begin{equation}
\begin{aligned}
\Pi^{(88)}_{0}\equiv  -\frac{4}{9}\pi \alpha_s [\langle \bar{u}u \rangle_0(\langle \bar{u}u \rangle_N+ \langle \bar{d}d \rangle_N)+4\langle \bar{s}s \rangle_0\langle \bar{s}s \rangle_N]\\
-\frac{5}{8}M^3(A^u_4+A^d_4+4A^s_4)+\frac{\pi^2}{36s}\langle(\frac{\alpha_s}{\pi}G^2)^2\rangle_N \big(\frac{13}{6}+ln\frac{s}{4}\big)\\
-\frac{4\pi}{s}\big[\frac{1}{27}\langle m_s\bar{s}s\alpha_sG^2\rangle_N +8\alpha_sm_s^2\langle (\bar{s}t^a\gamma_5s)(\bar{s}t^a\gamma_5s) \rangle_N \big]  \\  +\frac{51}{32}\frac{M^5}{s}(A^u_6+A^d_6+4A^s_6) \\
=0
\end{aligned}
\end{equation}
 From equating  coefficients of $(\vec{q}^2)^2$  we get :
\begin{equation}
\begin{aligned}
\Pi^{(88)}_{41}\equiv  -\frac{2}{9}\pi \alpha_s\frac{M}{s^2} [\langle \bar{u}u \rangle_0(\langle \bar{u}u \rangle_N+ \langle \bar{d}d \rangle_N)+4\langle \bar{s}s \rangle_0\langle \bar{s}s \rangle_N]\\
-\frac{61}{16}\frac{M^4}{s^2}(A^u_4+A^d_4+4A^s_4)+\frac{M\pi^2}{72s^3}\langle(\frac{\alpha_s}{\pi}G^2)^2\rangle_N \big(\frac{13}{6}+ln\frac{s}{4}\big)\\
-\frac{2\pi M}{s^3}\Big[\frac{1}{27}\langle m_s\bar{s}s\alpha_sG^2\rangle_N +8\alpha_sm_s^2\langle( \bar{s}t^a\gamma_5s)( \bar{s}t^a\gamma_5s) \rangle_N \Big]  \\  +\frac{1345}{192}\frac{M^6}{s^3}(A^u_6+A^d_6+4A^s_6) \\
=(\chi_8)^2+ 2M^2\sum_i(\chi_{8i})^2e^{-(M_i-M)^2/s}\Big[\frac{M}{2M_i^3}-\\
\frac{2M(M_i-M)}{M_i^2}\frac{\chi'_{8i}}{\chi_{8i}} -\frac{(M_i-M)^2}{sM_i^2}\Big]
\end{aligned}
\end{equation}
We can evaluate OPE expressions using following factorization  approximations and values of the parameters given in Table I :
\begin{eqnarray}
\langle m_s\bar{s}s\frac{\alpha_s}{\pi}G^2\rangle_N \approx m_s[\langle \bar{s}s \rangle_0\langle\frac{\alpha_s}{\pi}G^2\rangle_N+\langle \bar{s}s \rangle_N\langle\frac{\alpha_s}{\pi}G^2\rangle_0]\\
\langle(\frac{\alpha_s}{\pi}G^2)^2\rangle_N  \approx 2\langle\frac{\alpha_s}{\pi}G^2\rangle_0\langle\frac{\alpha_s}{\pi}G^2\rangle_N\\
\langle (\bar{s}t^a\gamma_5s)( \bar{s}t^a\gamma_5s) \rangle_N\approx  -\frac{2}{9}\langle \bar{s}s \rangle_0\langle \bar{s}s \rangle_N
\end{eqnarray}

\begin{table}[h!]
\begin{center}
\caption{{\footnotesize Numerical values of the QCD and phenomenological parameters  used  in \\ this work. All dimensional parameters have been expressed in GeV units. }} \label{table:1}
\begin {tabular}{ lc lc lc lc lc lc lc| }
\hline \hline
Parameter & $A^u_2$ & $A^d_2$ & $A^s_2$ & $A^u_4$ & $A^d_4$ & $A^s_4$ \\
\hline
Value(calculated) & 0.8074 & 0.4334 & 0.0479 & 0.1901 & 0.0304 & 0.0011 \\
\hline
Ref. & \cite{Martin} & \cite{Martin} & \cite{Martin} & \cite{Martin} & \cite{Martin} & \cite{Martin} \\
 \hline \hline
Parameter & $A^u_6$ & $A^d_6$ & $A^s_6$ &   &   &   \\
\hline
Value(calculated) & 0.0497 & 0.0055 & 0.0001 &   &    &   \\
\hline
Ref. & \cite{Martin} & \cite{Martin} & \cite{Martin} &   &   &   \\
\hline \hline
Parameter & $m_u\langle \bar{u}u \rangle_N$ & $m_u\langle \bar{u}u \rangle_N$ &  $m_s\langle \bar{s}s \rangle_N$& $\langle \bar{u}u \rangle_0$   & $\langle \bar{d}d \rangle_0$   & $\langle \bar{s}s \rangle_0$  \\
   & $+ m_d\langle \bar{d}d \rangle_N$ & $- m_d\langle \bar{d}d \rangle_N$ &
   &    &    &   \\  
\hline
Value(calculated) & 0.0554 & -0.0202 & 0.050 &  0.994$\langle \bar{q}q \rangle_0$  & 1.006$\langle \bar{q}q \rangle_0$    & 0.8$\langle \bar{q}q \rangle_0$  \\
\hline
Ref. & \cite{Hoferichter} & \cite{Hoferichter} & \cite{Copeland}  & \cite{Nicola}  & \cite{Nicola}  & \cite{Ioffeb}  \\
\hline \hline
Parameter & $m_u(1GeV)$ & $m_d(1GeV)$ & $m_s(1GeV)$ & $\langle \bar{q}q \rangle_0$ & $\langle\frac{\alpha_s}{\pi}G^2\rangle_0$ & $\alpha_s(1GeV)$ \\
\hline
Value & 0.0029 & 0.0071 & 0.147 & 0.017 & 0.012 & 0.4736 \\
\hline
Ref. & \cite{Ioffeb} & \cite{Ioffeb} & \cite{Ioffeb} & \cite{Ioffeb} & \cite{Ioffeb} & \cite{Tanaka} \\
 \hline \hline
\end{tabular}
\end{center}
\end{table}

\begin{figure}[h]
\centering
\includegraphics[width=0.8\linewidth]{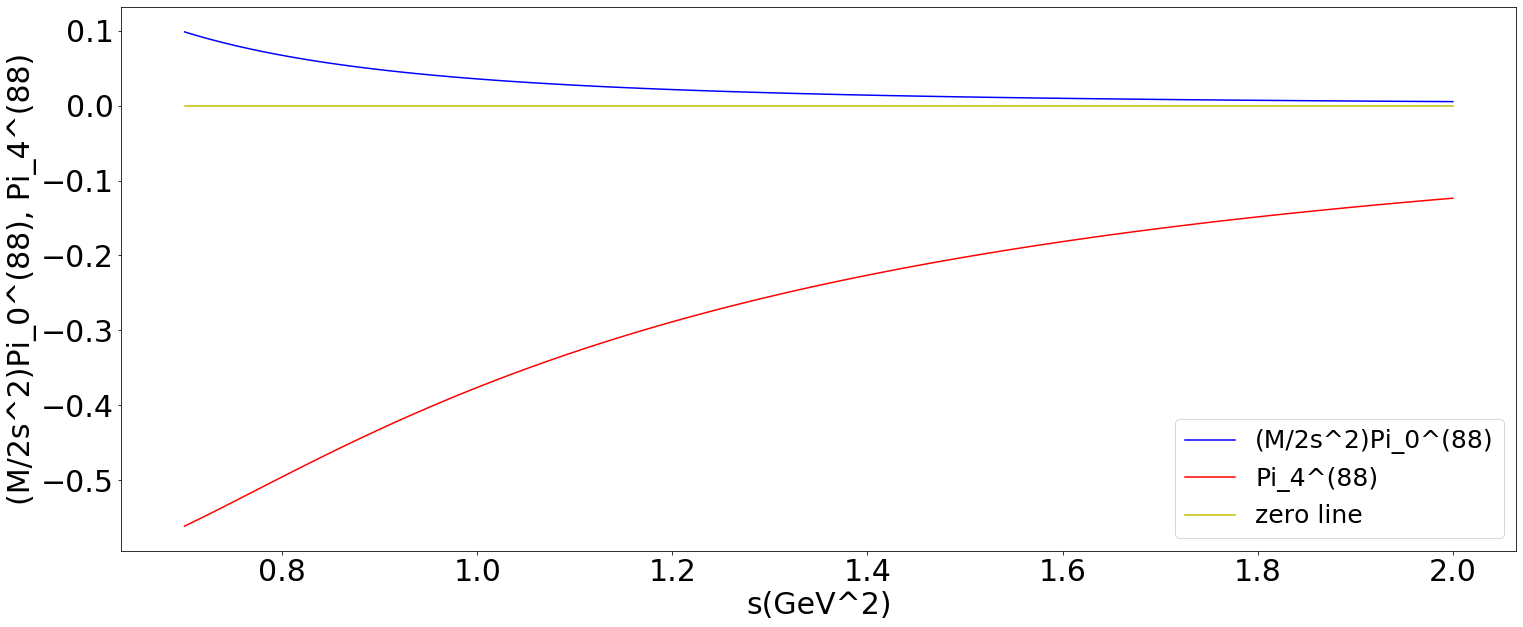}
\caption{{\scriptsize Plots of expressions of $\frac{M}{2s^2}\Pi_0^{(88)}$, $\Pi_4^{(88)}$ and a zero-line for comparision. }}
\label{fig:1}
\end{figure}

\begin{figure}[htbp]
\centering
\begin{subfigure}[b]{0.48\textwidth}
\centering
\includegraphics[width=\textwidth]{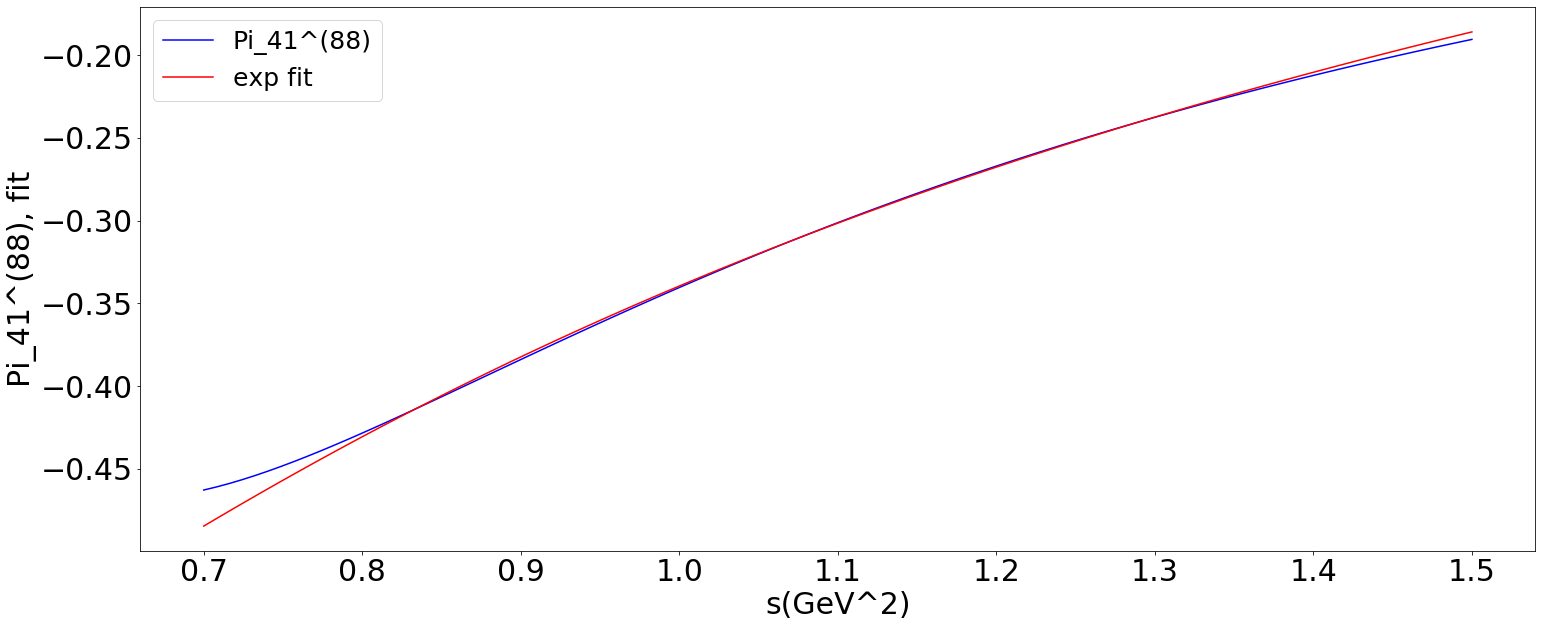}
\caption{{\scriptsize Plots of OPE expression of $\Pi_{41}^{(88)}$ and its fit  as  $0.258+(-0.827/s)e^{-0.325/s}$ as functions of s. }}
\label{fig : part1}
\end{subfigure}
\hfill
\begin{subfigure}[b]{0.48\textwidth}
\centering
\includegraphics[width=\textwidth]{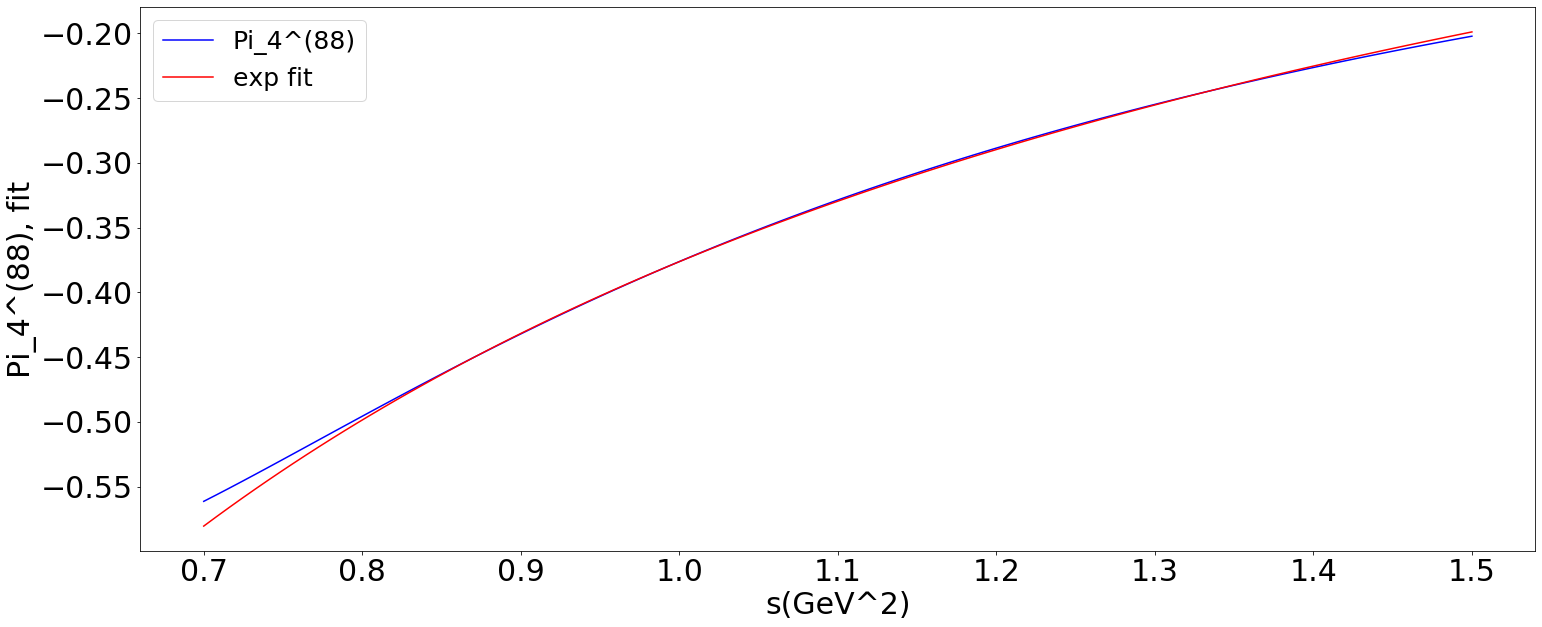}
\caption{{\scriptsize Plots of OPE expression of $\Pi_4^{(88)}$ and its fit  as  $0.21+(-0.671/s)e^{-0.135/s}$  as functions of s.}}
\label{fig : part2}
\end{subfigure}
\caption{ Two plots, each one of which gives $(\chi_8)^2$. } 
\label{fig : main figure}
\end{figure}

From combination of $\Pi^{(88)}_{0}$ and  $\Pi^{(88)}_{41}$  one gets :
\begin{equation}
\begin{aligned}
(\chi_8)^2+ 2M^2\sum_i(\chi_{8i})^2e^{-(M_i-M)^2/s}\Big[\frac{M}{2M_i^3}-\\
\frac{2M(M_i-M)}{M_i^2}\frac{\chi'_{8i}}{\chi_{8i}} -\frac{(M_i-M)^2}{sM_i^2}\Big]\\=  -\frac{7}{2}\frac{M^4}{s^2}(A^u_4+A^d_4+4A^s_4)+\frac{149}{24}\frac{M^6}{s^3}(A^u_6+A^d_6+4A^s_6)\equiv\Pi^{(88)}_{4}
\end{aligned}
\end{equation}
It is amazing that $\chi_8$ and $\chi_{8i}$ depend only  $A_n^q$'s after combining the two sum rules given by Eqs. (24) and (25). The coefficient of $(\vec{q}^2)$ does not give any useful information.\\
\par \hspace{12pt} In a similar way we can proceed for $\chi_g \chi_8$.

From equating  coefficients of $(\vec{q}^2)^0$ and $(\vec{q}^2)^2$  we get :
\begin{equation}
\begin{aligned}
\Pi^{(08)}_{0}\equiv-\frac{8}{9}\pi \alpha_s [\langle \bar{u}u \rangle_0(\langle \bar{u}u \rangle_N+ \langle \bar{d}d \rangle_N)-4\langle \bar{s}s \rangle_0\langle \bar{s}s \rangle_N]\\
-\frac{5}{4}M^3(A^u_4+A^d_4-4A^s_4)+8m_s^2\big[2m_s\langle \bar{s}s \rangle_N+\frac{1}{8}\langle\frac{\alpha_s}{\pi}G^2\rangle_N-\frac{3}{2}MA^s_2+\\     \frac{176}{27}\frac{\pi \alpha_s}{s}\langle \bar{s}s \rangle_0\langle \bar{s}s \rangle_N +\frac{5}{2}\frac{M^3}{s}A^s_4\big]+  \frac{\pi^2}{9s}\big[-\langle(\frac{\alpha_s}{\pi}G^2)^2\rangle_N \frac{1}{6}(\frac{13}{6}+ln\frac{s}{4}\big)\\
+\frac{8}{3}\langle m_s\bar{s}s\frac{\alpha_s}{\pi}G^2\rangle_N\big]  +\frac{51}{16}\frac{M^5}{s}(A^u_6+A^d_6-4A^s_6) =0
\end{aligned}
\end{equation}

\begin{equation}
\begin{aligned}
\Pi^{(08)}_{41}\equiv-\frac{4}{9}\pi \alpha_s\frac{M}{s^2} [\langle \bar{u}u \rangle_0(\langle \bar{u}u \rangle_N+ \langle \bar{d}d \rangle_N)-4\langle \bar{s}s \rangle_0\langle \bar{s}s \rangle_N]\\
-\frac{61}{8}\frac{M^4}{s^2}(A^u_4+A^d_4-4A^s_4)+\frac{4Mm_s^2}{s^2}\Big[2m_s\langle \bar{s}s \rangle_N \\
+\frac{1}{8}\langle\frac{\alpha_s}{\pi}G^2\rangle_N-\frac{11}{2}MA^s_2+   
\frac{176}{27}\frac{\pi \alpha_s}{s}\langle \bar{s}s \rangle_0\langle \bar{s}s \rangle_N \\ +\frac{91}{6}\frac{M^3}{s}A^s_4\Big]  +\frac{M\pi^2}{18s^3}\Big[-\langle(\frac{\alpha_s}{\pi}G^2)^2\rangle_N \frac{1}{6}\big(\frac{13}{6}+ln\frac{s}{4}\big) \\ +\frac{8}{3}\langle m_s\bar{s}s\frac{\alpha_s}{\pi}G^2\rangle_N\Big]  
 +\frac{1345}{96}\frac{M^6}{s^3}(A^u_6+A^d_6-4A^s_6) \\
=\chi_g\chi_8+ 2M^2\sum_i\chi_{gi}\chi_{8i}e^{-(M_i-M)^2/s}\Big[\frac{M}{2M_i^3}-\\
\frac{M(M_i-M)}{M_i^2}\big(\frac{\chi'_{gi}}{\chi_{gi}}+\frac{\chi'_{8i}}{\chi_{8i}}\big) -\frac{(M_i-M)^2}{sM_i^2}\Big]
\end{aligned}
\end{equation}

\begin{figure}[h]
\centering
\includegraphics[width=0.8\linewidth]{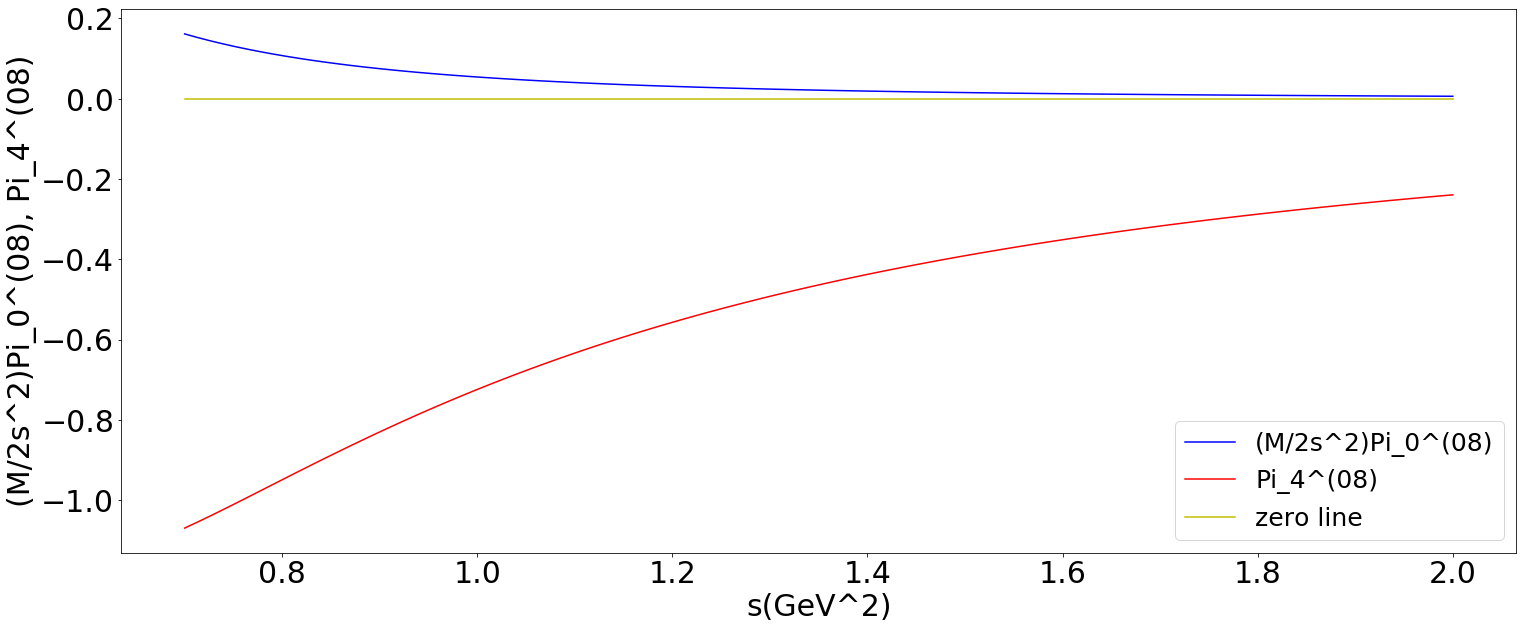}
\caption{{\scriptsize Plots of expressions of $\frac{M}{2s^2}\Pi_0^{(08)}$, $\Pi_4^{(08)}$ and a zero-line for comparision. }}
\label{fig:4}
\end{figure}

From combination of $\Pi^{(08)}_{0}$ and  $\Pi^{(08)}_{41}$ one gets :
\begin{equation}
\begin{aligned}
\Pi^{(08)}_{4}\equiv -7\frac{M^4}{s^2}(A^u_4+A^d_4-4A^s_4)-\frac{16m_s^2M^2}{s^2}A^s_2+\\    \frac{152}{3}\frac{m_s^2M^4}{s^3}A^s_4+ \frac{149}{12}\frac{M^6}{s^3}(A^u_6+A^d_6-4A^s_6) \\
=\chi_g\chi_8+ 2M^2\sum_i\chi_{gi}\chi_{8i}e^{-(M_i-M)^2/s}\Big[\frac{M}{2M_i^3}-\\
\frac{M(M_i-M)}{M_i^2}\big(\frac{\chi'_{gi}}{\chi_{gi}}+\frac{\chi'_{8i}}{\chi_{8i}}\big) -\frac{(M_i-M)^2}{sM_i^2}\Big]\\
\end{aligned}
\end{equation}

\begin{figure}[htbp]
\centering
\begin{subfigure}[b]{0.48\textwidth}
\centering
\includegraphics[width=\textwidth]{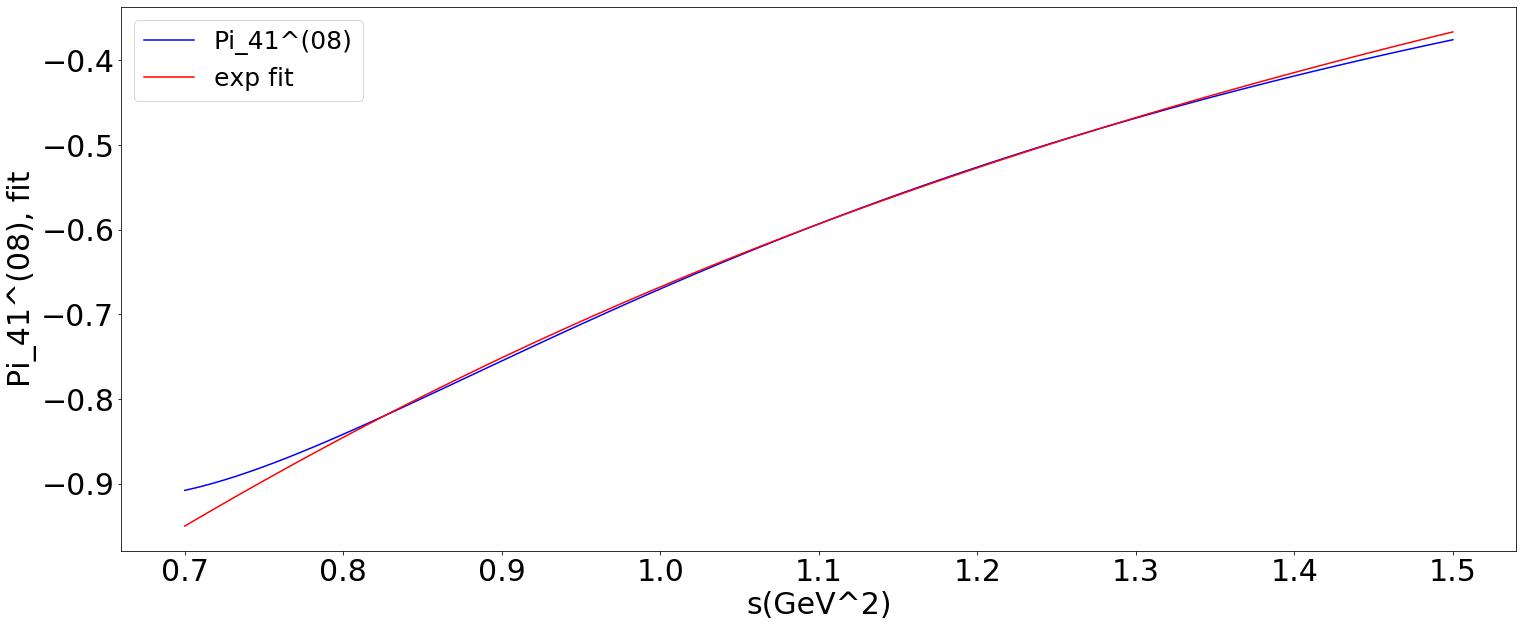}
\caption{{\scriptsize Plots of OPE expression of $\Pi_{41}^{(08)}$ and its fit  as  $0.511+(-1.641/s)e^{-0.331/s}$ as  functions of s. }}
\label{fig : part1}
\end{subfigure}
\hfill
\begin{subfigure}[b]{0.48\textwidth}
\centering
\includegraphics[width=\textwidth]{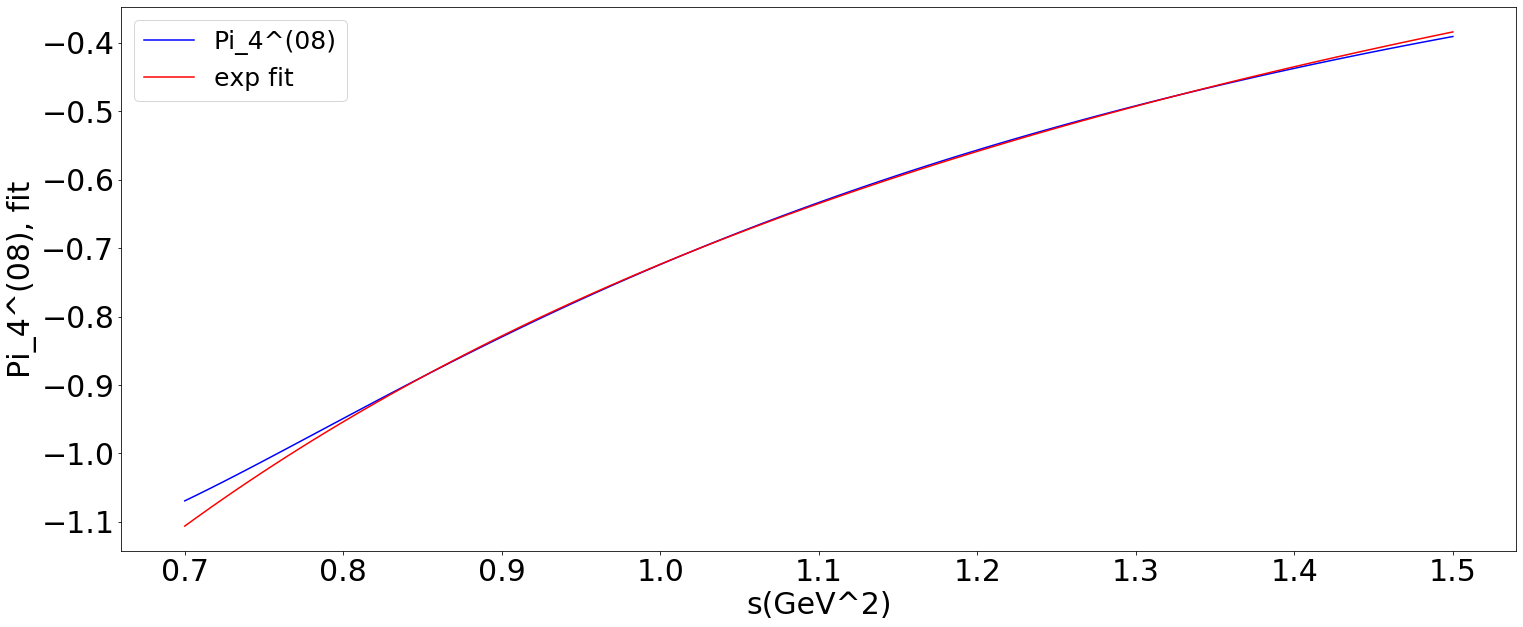}
\caption{{\scriptsize Plots of OPE expression of $\Pi_4^{(08)}$ and its fit  as  $0.418+(-1.337/s)e^{-0.158/s}$ as  functions of s.}}
\label{fig : part2}
\end{subfigure}
\caption{ Two plots, each one of which gives $\chi_g\chi_8$. } 
\label{fig : main figure}
\end{figure}
 
In Figs. (1, 3), we have shown that $(M/(2s^2))|\Pi^{(88)}_{0}|<<|\Pi^{(88)}_{4}|$ and $(M/(2s^2))|\Pi^{08)}_{0}|<<|\Pi^{(08)}_{4}|$. In principle, $\Pi^{(88)}_{41}$  or $\Pi^{(88)}_{4}$ should be fitted as $a+(b+c/s)e^{d/s}$ with the requirement that $ a>0$ and $c,d<0$, whereas there is a priori no requirement on b. However, such  functions can be parameterized in terms of just 3 parameters and the fourth is redundant. We choose these three to be a, c and d. Also, $a+(b)e^{d/s}$ parameterization does not work.   Among the  $\Pi^{(88)}_{41}$ and $\Pi^{(88)}_{4}$ it is preferable to choose  $\Pi^{(88)}_{4}$, since it has minimum number of parameters reducing the uncertainty. On the other hand for $\Pi^{(08)}_{41}$ and  $\Pi^{(08)}_{4}$ there is no such requirement for signs of $a$ and $c$ though we would like $d<0$. Using QCD trace anomaly \cite{Collins} we can write :
$\langle\frac{\alpha_s}{\pi}G^2\rangle_N=-\frac{8}{9}[M-(m_u\langle \bar{u}u\rangle_N+m_d\langle \bar{d}d \rangle_N +m_s\langle \bar{s}s \rangle_N)] $.
Fitting is done over 0.8 $ GeV^2<s<1.5$ $GeV^2$.  From Fig.2a (for $\Pi^{88}_{41}$) : $(\chi_8)^2=0.258$,  whereas from Fig.2b (for $\Pi^{88}_4$) :   $(\chi_8)^2=0.21$. From Fig.4a (for $\Pi^{08}_{41}$) : $\chi_g\chi_8=0.511$ and from Fig.4b (for $\Pi^{08}_4$) : $\chi_g\chi_8=0.418$. If the range of fitting is varied, then the numbers will change.\\
\par \hspace{10pt}From $(\chi_8)^2=0.21, 0.258$ and $\chi_g\chi_8=0.418, 0.511$, one gets : \\
\begin {equation} |\chi_8|=0.483\pm 0.025, \; |\chi_g|=1.014\pm 0.102. \end {equation}

\section{Isovector and isoscalar axial currents}
In this section we consider nuclean matrix elements of correlators of  axial-vector currents. For calculating $g^3_A$ and $g^{u+d}_A$, we introduce the currents

\begin {equation}
\begin{aligned}
J^{3}_{\mu5}=\frac{1}{2}\eta_{\mu \nu}(\bar{u}\gamma^{\nu}\gamma_5 u-\bar{d}\gamma^{\nu}\gamma_5 d)\\
\eta_{\mu \nu}=\frac{q_{\mu}q_{\nu}}{q^2}-g_{\mu \nu} \\
J^{u+d}_{\mu5}= \bar{u}\gamma_{\mu}\gamma_5 u+\bar{d}\gamma_{\mu}\gamma_5 d\\
\end{aligned}
\end{equation}
$\eta_{\mu \nu}$ is used to make the current conserved and to suppress pion  contribution to the current \cite{Nishikawa, Reinders}. Consider the correlators
\begin{equation}
\begin{aligned}
\Pi^{(33)}_{\mu\nu}(q^2) \equiv  i \int d^4xe^{iqx}\langle T\{J^{3}_{\mu5}(x), J^{3}_{\nu5}(0)  \}\rangle_N \\
\Pi^{(03)}_{\mu\nu}(q^2) \equiv  i \int d^4xe^{iqx}\langle T\{J^{u+d}_{\mu5}(x), J^{3}_{\nu5}(0)  \}\rangle_N \\
\end{aligned}
\end{equation}
and the form factors
\begin {equation}
\begin{aligned}
\langle P,S|J^3_{\mu5}|P+q,S' \rangle = \frac{\eta_{\mu \nu}}{2}\bar{u}(P,S)g^3_A(q^2)\lambda^3\gamma^{\nu}\gamma_5 u(P+q,S')\\
\langle P,S|J^{u+d}_{\mu5}|P+q,S' \rangle = \bar{u}(P,S)[g^{u+d}_A(q^2)\gamma_{\mu}\gamma_5+h^{u+d}_A(q^2)q_{\mu}\gamma_5] u(P+q,S').
\end{aligned}
\end{equation}
Call $\Pi^{(ij)\mu}_{\mu}=\Pi^{(ij)}$.

\begin{equation}
\begin{aligned}
\Pi^{(33)}_{OPE}(q, P)=\frac{1}{4}\Big[\frac{6}{q^2}(m_u\langle \bar{u}u \rangle_N+ m_d\langle \bar{d}d \rangle_N)-\frac{1}{2q^2}\langle\frac{\alpha_s}{\pi}G^2\rangle_N\\    -8i\frac{q^{\mu}q^{\nu}}{q^4}(\langle \bar{u}S\gamma_{\mu}D_{\nu}u\rangle_N +\langle  \bar{d}S\gamma_{\mu}D_{\nu}d\rangle_N)\\  - \frac{7}{6}\frac{\pi\alpha_s}{q^4}\langle S (\bar{u}\gamma_{\mu}\lambda^au + \bar{d}\gamma_{\mu}\lambda^ad)\sum_q\bar{\psi}_q\gamma^{\mu}\lambda^a\psi_q\rangle_N\\  + 2\pi\alpha_s\frac{q^{\mu}q^{\nu}}{q^6}\langle S(\bar{u}\gamma_{\mu}\lambda^au + \bar{d}\gamma_{\mu}\lambda^ad)\sum_q\bar{\psi}_q\gamma_{\nu}\lambda^a\psi_q\rangle_N\\   + 8\frac{\pi\alpha_s}{q^4} \Big\{\frac{q^{\mu}q^{\nu}}{q^2}\langle S(\bar{u}\gamma_{\mu}\lambda^au - \bar{d}\gamma_{\mu}\lambda^ad)(\mu\rightarrow \nu)\rangle_N \\- \langle (\bar{u}\gamma_{\mu}\lambda^au - \bar{d}\gamma_{\mu}\lambda^ad)^2\rangle_N\Big\}\\
+48i\frac{q^{\alpha}q^{\beta}q^{\lambda}q^{\sigma}}{q^8}(\langle \bar{u}S \gamma_{\sigma}D_{\alpha}D_{\beta}D_{\lambda}u\rangle_N +\langle \bar{d}S\gamma_{\sigma}D_{\alpha}D_{\beta}D_{\lambda}d\rangle_N)\Big]
\end{aligned}
\end{equation}
Using factorization approximation and expressing nucleon matrix elements containing covariant derivatives in terms of moments of parton distribution function, we can express  $\Pi^{(33)}_{OPE}(q, P)$ as 

\begin {equation}
\begin{aligned}
\Pi^{(33)}_{OPE}(\omega, \vec{q})=\frac{1}{4}\Big[\frac{6}{q^2}(m_u\langle \bar{u}u \rangle_N+ m_d\langle \bar{d}d \rangle_N) -\frac{1}{2}\frac{1}{q^2}\langle\frac{\alpha_s}{\pi}G^2\rangle_N\\
-4M(A^u_2+A^d_2)\big(\frac{\omega^2}{q^4}-\frac{1}{4}\frac{1}{q^2}\big)\\
+\frac{640}{27}\frac{\pi \alpha_s}{q^4}\langle \bar{u}u \rangle_0(\langle \bar{u}u \rangle_N+ \langle \bar{d}d \rangle_N)\\
-24M^3(A^u_4+A^d_4)\big(\frac{\omega^4}{q^8}-\frac{3}{4}\frac{\omega^2}{q^6}+\frac{1}{16q^4}\big)\Big]
\end{aligned}
\end{equation}
For phenomenological expression we write $\Pi^{(33)}_{ph}(\omega, \vec{q})$ using spectral function
\begin{equation}
\begin{aligned}
\Pi^{(33)}_{ph}(\omega, \vec{q})=\int_{-\infty}^{+\infty}d\omega'\frac{\rho_{33}(\omega',\vec{q})}{\omega-\omega'},\\
B\Big[\Pi^{(33)}_{ph}(\omega, \vec{q})\Big]=-\frac{\vec{q}^2}{2M}(g^3_A)^2e^{-\vec{q}^4/(4Ms)}-\sum_i\frac{M}{2M_i}(g^3_{Ai})^2\big\{6\Delta M_i e^{-(\Delta M_i)^2/s}\\ +\vec{q}^2\Big[e^{-(\Delta M_i)^2/s}
\Big(\frac{1}{M_i}-\frac{3}{2}\frac{\Delta M_i}{M_i^2}+\frac{2}{\Delta M_i}-\frac{6(\Delta M_i)^2}{sM_i}\Big)\Big]+O(\vec{q}^4)\big\}
\end{aligned}
\end{equation}
where $\rho_{33}$ denotes spectral function and  $\Delta M_i=M_i-M$. On taking first derivative with respect to $\vec{q}^2$  of the Borel transform of $\Pi^{(33)}(\omega, \vec{q})$ we get
\begin{equation}
\begin{aligned}
\frac{\partial}{\partial \vec{q}^2}B\big[\Pi^{(33)}_{ph}\big]_{\vec{q}^2=0}=-\frac{1}{2M}\Big\{(g^3_A)^2+\sum_i (g^3_{Ai})^2e^{-(\Delta M_i)^2/s}\frac{M^2}{M_i}\\ \times \Big[\frac{1}{M_i}-\frac{3}{2}\frac{\Delta M_i}{M_i^2}+\frac{2}{\Delta M_i}-\frac{6(\Delta M_i)^2}{sM_i}\Big]\Big\}
\end{aligned}
\end{equation}

\begin{equation}
\begin{aligned}
\frac{\partial}{\partial \vec{q}^2}B\big[\Pi^{(33)}_{OPE}\big]_{\vec{q}^2=0}=\frac{1}{4}\Big[\frac{6}{s}(m_u\langle \bar{u}u \rangle_N+ m_d\langle \bar{d}d \rangle_N) -\frac{1}{2s}\langle\frac{\alpha_s}{\pi}G^2\rangle_N\\  -\frac{7M}{s}(A^u_2+A^d_2) -\frac{640}{27}\frac{\pi \alpha_s}{s^2}\langle \bar{u}u \rangle_0(\langle \bar{u}u \rangle_N+ \langle \bar{d}d \rangle_N)\\
+\frac{45}{2}\frac{M^3}{s^2}(A^u_4+A^d_4)\Big]
\end{aligned}
\end{equation}
Similar steps are taken for $\Pi^{(03)}(\omega, \vec{q})$ :
\begin{equation}
\begin{aligned}
\frac{\partial}{\partial \vec{q}^2}B\big[\Pi^{(03)}_{ph}\big]_{\vec{q}^2=0}=\frac{1}{M}g^{u+d}_Ag^3_A+\sum_ig^{u+d}_{Ai} g^3_{Ai} e^{-(\Delta M_i)^2/s}\frac{M \Delta M_i}{M_i^3}\\  \Big[\frac{2MM_i}{(\Delta M_i)^2}-\frac{3}{2}+\frac{3M_i}{\Delta M_i}-\frac{6 M_i\Delta M_i}{s}\Big]
\end{aligned}
\end{equation}
\begin{equation}
\begin{aligned}
\frac{\partial}{\partial \vec{q}^2}B\big[\Pi^{(03)}_{OPE}\big]_{\vec{q}^2=0}=-\frac{3}{s}(m_u\langle \bar{u}u \rangle_N- m_d\langle \bar{d}d \rangle_N)\\
+\frac{7M}{2s}(A^u_2-A^d_2)
+\frac{320}{27}\frac{\pi \alpha_s}{s^2}(\langle \bar{u}u \rangle_0\langle \bar{u}u \rangle_N-\langle \bar{d}d \rangle_0 \langle \bar{d}d \rangle_N)\\
-\frac{45}{4}\frac{M^3}{s^2}(A^u_4-A^d_4)
\end{aligned}
\end{equation}

\begin{figure}[h]
\centering
\includegraphics[width=0.8\linewidth]{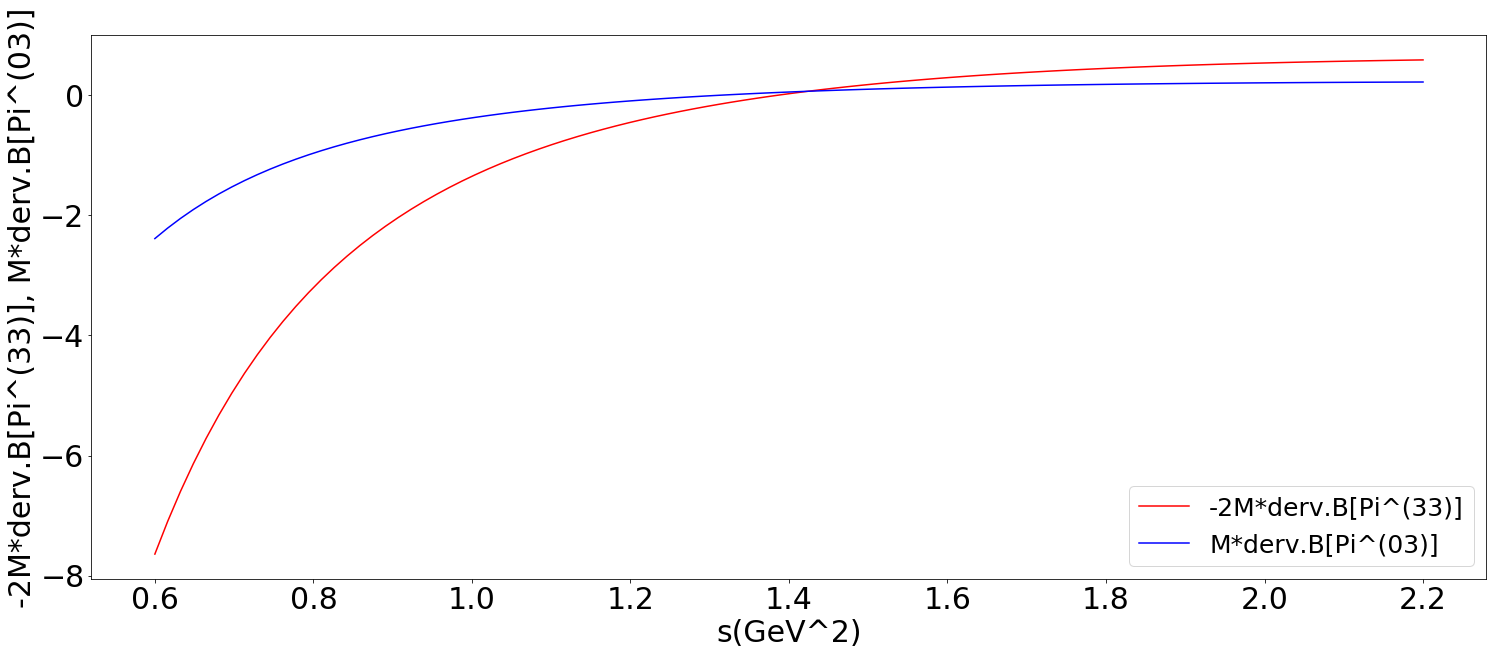}
\caption{{\scriptsize Plots of OPE expressions of  -2M$\frac{\partial}{\partial \vec{q}^2}B\big[\Pi^{(33)}\big]_{\vec{q}^2=0}$ and M$\frac{\partial}{\partial \vec{q}^2}B\big[\Pi^{(03)}\big]_{\vec{q}^2=0}$  \\as  functions of s. }}
\label{fig:1}
\end{figure}

\begin{figure}[h]
\centering
\includegraphics[width=0.8\linewidth]{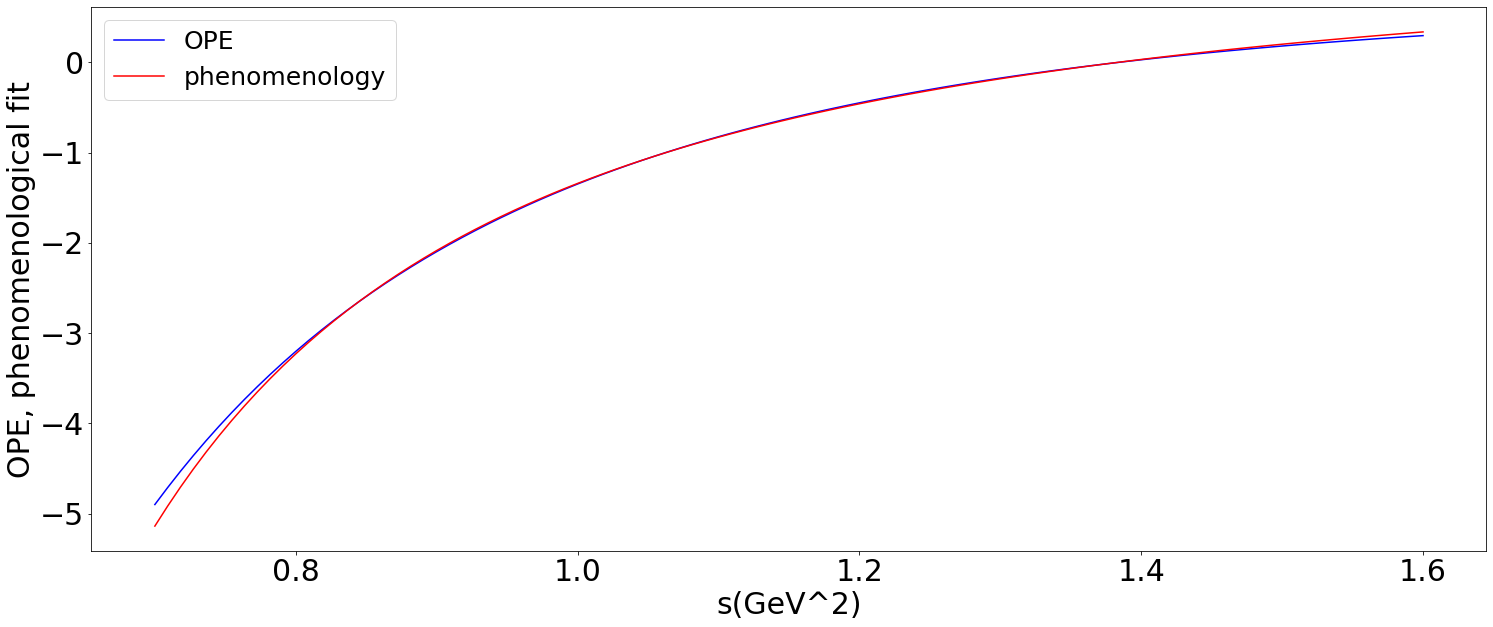}
\caption{{\scriptsize Plots of OPE expression of -2M$\frac{\partial}{\partial \vec{q}^2}B\big[\Pi^{(33)}\big]_{\vec{q}^2=0}$ and its fit  as  $1.479-  (0.887/s)  \\ \times e^{1.157/s}$ as functions of s. }}
\label{fig:2}
\end{figure}

\begin{figure}[h]
\centering
\includegraphics[width=0.8\linewidth]{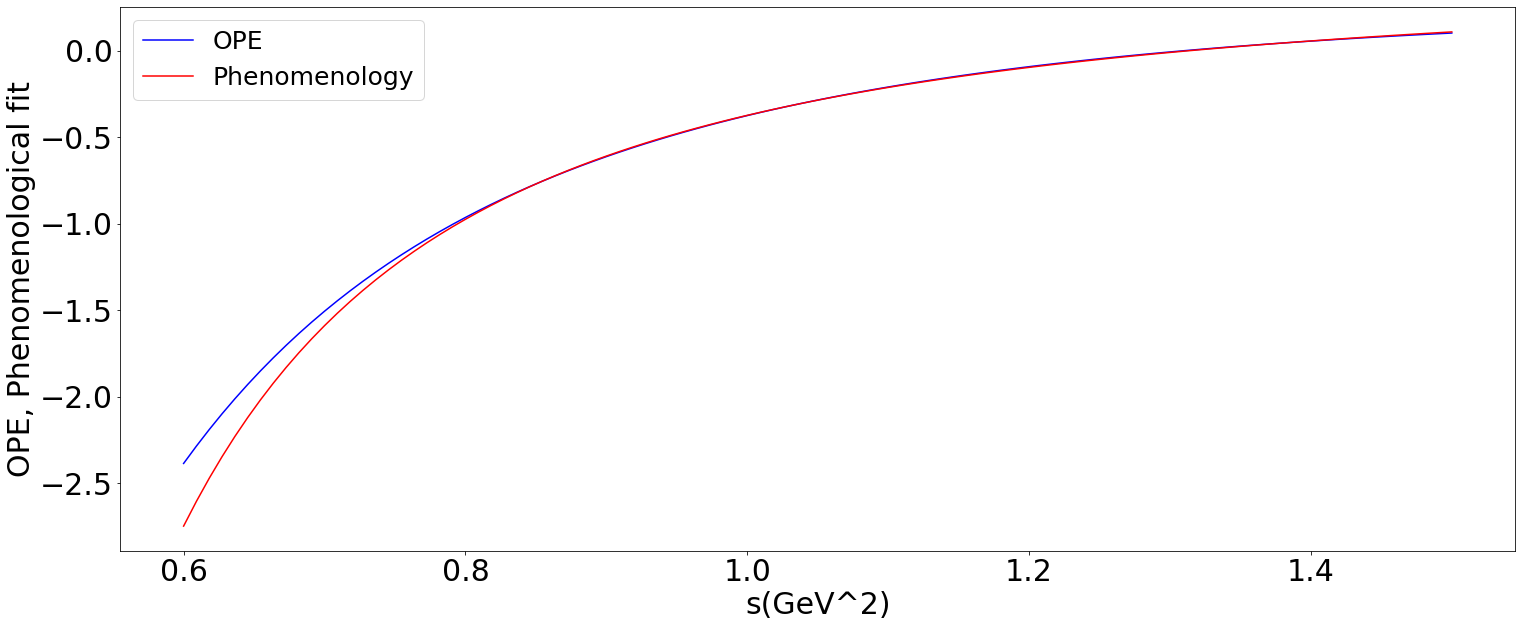}
\caption{{\scriptsize Plots of OPE expression of  M$\frac{\partial}{\partial \vec{q}^2}B\big[\Pi^{(03)}\big]_{\vec{q}^2=0}$ and its fit  as  $0.497-(0.26/s) \\\times e^{1.208/s}$ as functions of s.}}
\label{fig:3}
\end{figure}

Notice the similarity between Eqs. (41) and (43). From Figs. 6 and 7 we get $(g^3_A)^2=1.479$ and  $g^3_A g_A^{u+d}=0.497$. This gives $|g^3_A|=1.216$ and  $|g^{u+d}_A|=0.409$.\\
\par \hspace{12pt} On a similar line we proceed for the s-quark axial current:
\begin {equation}
\begin{aligned}
J^{s}_{\mu5}= \bar{s}\gamma_{\mu}\gamma_5 s\\
\langle P,S|J^s_{\mu5}(0)|P+q,S' \rangle = \bar{u}(P,S)[g^s_A(q^2)\gamma_{\mu}\gamma_5+h^s_A(q^2)q_{\mu}\gamma_5] u(P+q,S') \\
\Pi^{(ss)}_{\mu\nu}(q^2) \equiv  i \int d^4xe^{iqx}\langle T\{J^{s}_{\mu5}(x), J^{s}_{\nu5}(0)  \}\rangle_N
\end{aligned}
\end{equation}
On matching OPE and phenomenological expressions of $\Pi^{(ss)}(q^2)$, one gets
\begin {equation}
\begin{aligned}
(g^s_A)^2+\sum_i (g^s_{Ai})^2e^{-(\Delta M_i)^2/s}\frac{M^2}{M_i^2}\Big[1-\frac{2}{3}\frac{\Delta M_i}{M_i}-\frac{2(\Delta M_i)^2}{s}\Big]\\
=-\frac{M}{3}\Big[\frac{6}{s}m_s\langle \bar{s}s \rangle_N -\frac{1}{4s}\langle\frac{\alpha_s}{\pi}G^2\rangle_N
-\frac{7M}{s}A^s_2\\
-\frac{704}{27}\frac{\pi \alpha_s}{s^2}\langle \bar{s}s \rangle_0\langle \bar{s}s \rangle_N
+15\frac{M^3}{s^2}A^s_4\Big]
\end{aligned}
\end{equation}

\begin{figure}[h]
\centering
\includegraphics[width=0.8\linewidth]{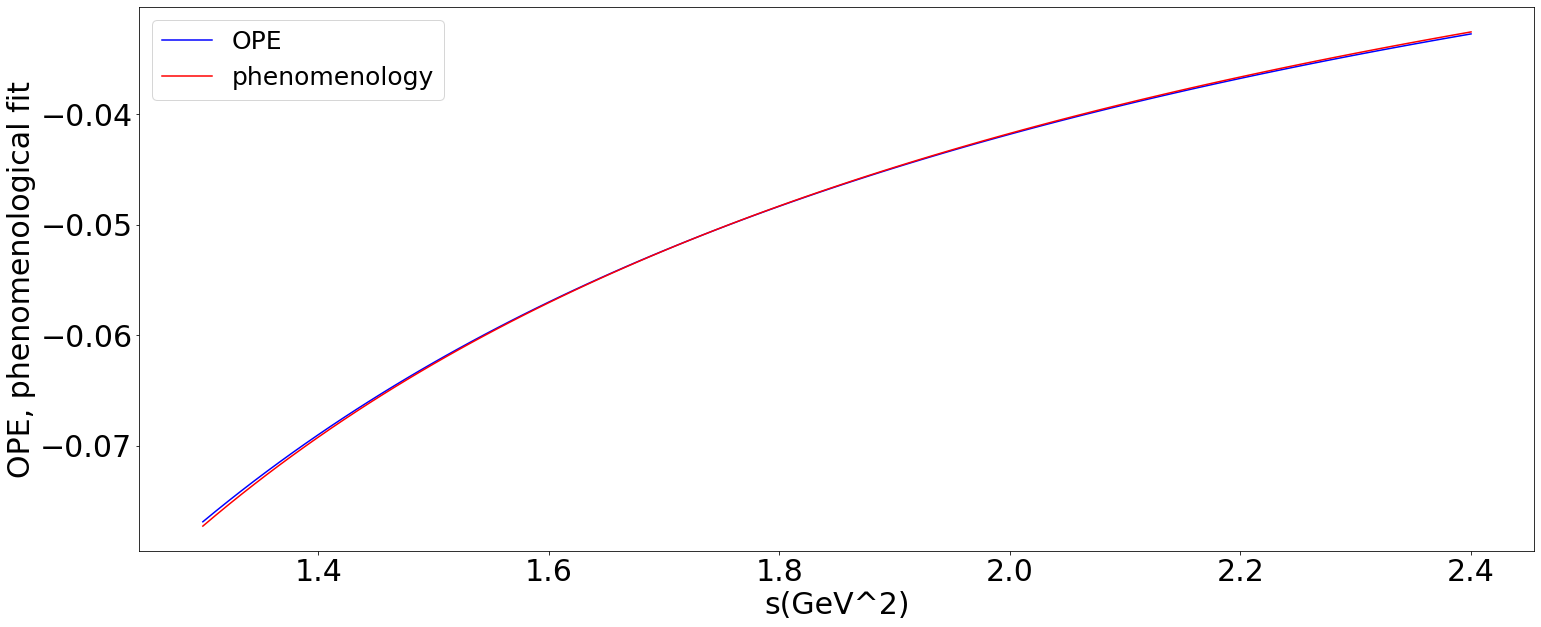}
\caption{{\scriptsize Plots of OPE expression of $-\frac{M}{3}\frac{\partial}{\partial \vec{q}^2}B\big[\Pi^{(ss)}\big]_{\vec{q}^2=0}$ and its fit  as  $0.003-(0.067/s) \\ \times e^{0.574/s}$ as functions of s.}}
\label{fig:5}
\end{figure}
From Fig. 8 we get $(g_A^s)^2=0.003$ which gives $|g_A^s|=0.054$.
A drawback of this determination of $g_A^s$ is that it does not take into account chiral anomaly and the resulting instanton contribution which is known to play significant role in this framework when singlet-singlet correlation is used to calculate $g_A^0$ \cite{Nishikawa,  Nishikawa1}. A way to tackle this problem has been suggested and implemented in Ref. \cite{JPS1} and we have applied it above for calculating  $|g_A^{u+d}|$  by considering a singlet-triplet axial current correlator along with triplet-triplet axial current correlator. However, for  $g_A^s$ we do not have the corresponding axial current for constructing a hybrid correlator. Also, note that the range of s over which OPE  and phenomenological expressions for $\Pi^{(ss)}(q^2)$ are matched is higher than the previous cases due to larger masses of the intermediate states. In previous works \cite{Nishikawa, Nishikawa1, JPS1} the continuum contribution was not taken into account. Unlike the cases of pseudoscalar coupling constants, for the axial coupling constants accounting for the continuum contribution requires positive exponent. One should remember that  a phenomenological  expression is parameterized as an an effective theory applicable in a limited range of s. \\
  
\section{Isovector and isoscalar pseudoscalar currents}
Here we will calculate nuclean matrix elements of pseudoscalar currents through  axial current-current correlator as done for octet pseudoscalar current. In the following we are introducing new notations for the sake of convenience : $J^+_{\mu 5}\equiv J^{u+d}_{\mu 5},$ but $ J^-_{\mu 5}$ is a modified form of $J^3_{\mu 5}$:

\begin {equation}
\begin{aligned}
J^{\pm}_{\mu 5}=\bar{u}\gamma_{\mu}\gamma_5 u\pm \bar{d}\gamma_{\mu}\gamma_5 d;\\
J^{\pm}_5=m_u\bar{u}i\gamma_5 u \pm m_d\bar{d}i\gamma_5 d, Q=\frac{\alpha_s}{\pi}G\tilde{G};\\
\partial^{\mu}J^-_{\mu5}=2J^-_5, \partial^{\mu}J^+_{\mu5}=2J^+_5+\frac{1}{2}Q 
\end{aligned}
\end{equation}

\begin {equation}
\begin{aligned}
\Pi_P^{--}(q^2) \equiv  i \int d^4xe^{iqx}\langle T\{J^-_5(x),J^-_5(0)\}\rangle_N \\
=\frac{i}{4}q^{\mu}q^{\nu} \int d^4xe^{iqx}\langle T\{J^-_{\mu5}(x),J^-_{\nu5}(0)\}\rangle_N -m_u\langle \bar{u}u\rangle_N-m_d\langle \bar{d}d\rangle_N
\end{aligned}
\end{equation}

\begin {equation}
\begin{aligned}
\Pi_P^{+-}(q^2) \equiv  i \int d^4xe^{iqx}\big[4\langle T\{J^+_5(x),J^-_5(0)\}\rangle_N +\langle T\{Q(x),J^-_5(0)\}\rangle_N\big]\\
= iq^{\mu}q^{\nu} \int d^4xe^{iqx}\langle T\{J^+_{\mu5}(x),J^-_{\nu5}(0)\}\rangle_N -4m_u\langle \bar{u}u\rangle_N+4m_d\langle \bar{d}d\rangle_N
\end{aligned}
\end{equation}

\begin {equation}\langle P,S|J^{\pm}_5(0)|P+q,S' \rangle = \chi^{\pm}_P(q^2)2M\bar{u}(P,S)i\gamma_5 u(P+q,S') \end{equation}

Following the same steps as for $\chi_8$ and $\chi_g$, we get
\begin{equation}
\begin{aligned}
(\chi^-_P)^2+ 2M^2\sum_i(\chi^-_{Pi})^2e^{-(M_i-M)^2/s}\Big[\frac{M}{2M_i^3}\\ -\frac{2M(M_i-M)}{M_i^2}\frac{\chi^{'-}_{Pi}}{\chi^-_{Pi}} -\frac{(M_i-M)^2}{sM_i^2}\Big]\\ =  -\frac{2}{9}\frac{\pi \alpha_sM}{s^2} [\langle \bar{u}u \rangle_0\langle \bar{u}u \rangle_N+ \langle \bar{d}d \rangle_0\langle \bar{d}d \rangle_N]-\frac{61}{16}\frac{M^4}{s^2}(A^u_4+A^d_4)\\
+\frac{M\pi^2}{216s^3}\langle \big(\frac{\alpha_s}{\pi}G^2\big)^2\rangle_N \Big(\frac{13}{6}+ln\frac{s}{4}\Big) +\frac{1345}{192}\frac{M^6}{s^3}(A^u_6+A^d_6) \\   -\frac{\pi^2 M}{s^3}\frac{1}{54}\big(\langle m_u\bar{u}u\frac{\alpha_s}{\pi}G^2\rangle_N +\langle m_d\bar{d}d\frac{\alpha_s}{\pi}G^2\rangle_N\big)  \equiv \Pi^{u-d}_{P41}  .
\end{aligned}
\end{equation}
Also,  as done for $\Pi^{88}$ and $\Pi^{08}$, we can deduce
\begin{equation}
\begin{aligned}
(\chi^-_P)^2+ 2M^2\sum_i(\chi^-_{Pi})^2e^{-(M_i-M)^2/s}\Big[\frac{M}{2M_i^3}\\ -\frac{2M(M_i-M)}{M_i^2}\frac{\chi^{'-}_{Pi}}{\chi^-_{Pi}} -\frac{(M_i-M)^2}{sM_i^2}\Big]\\ =  -\frac{7}{2}\frac{M^4}{s^2}(A^u_4+A^d_4)
 +\frac{149}{24}\frac{M^6}{s^3}(A^u_6+A^d_6)  \equiv \Pi^{u-d}_{P4}  .
\end{aligned}
\end{equation}
Next, from the correlator $\Pi_P^{+-}$ we get:
\begin{equation}
\begin{aligned}
\chi^-_P(\chi^+_P+\frac{1}{4}\chi_g)+ 2M^2\sum_i\chi^+_{Pi}\chi^-_{Pi}e^{-(M_i-M)^2/s}\Big[\frac{M}{2M_i^3}\\ -\frac{M(M_i-M)}{M_i^2}\Big(\frac{\chi^{'+}_{Pi}}{\chi^+_{Pi}}+\frac{\chi^{'-}_{Pi}}{\chi^-_{Pi}}\Big) -\frac{(M_i-M)^2}{sM_i^2}\Big]\\
+ 2M^2\sum_i\chi_{gi}\chi^-_{Pi}e^{-(M_i-M)^2/s}\Big[\frac{M}{8M_i^3} -\frac{M(M_i-M)}{4M_i^2}\Big(\frac{\chi'_{gi}}{\chi_{gi}}+\frac{\chi^{'-}_{Pi}}{\chi^-_{Pi}}\Big) -\frac{(M_i-M)^2}{4sM_i^2}\Big]\\
=  -\frac{2}{9}\frac{\pi \alpha_sM}{s^2} [\langle \bar{u}u \rangle_0\langle \bar{u}u \rangle_N- \langle \bar{d}d \rangle_0\langle \bar{d}d \rangle_N]-\frac{61}{16}\frac{M^4}{s^2}(A^u_4-A^d_4)\\
 -\frac{\pi^2 M}{s^3}\frac{1}{54}\big(\langle m_u\bar{u}u\frac{\alpha_s}{\pi}G^2\rangle_N -\langle m_d\bar{d}d\frac{\alpha_s}{\pi}G^2\rangle_N\big) \\
+\frac{1345}{192}\frac{M^6}{s^3}(A^u_6-A^d_6)  \equiv \Pi^{u+d}_{P41}  .
\end{aligned}
\end{equation}
Also,  as done for $\Pi^{--}$ above we can deduce
\begin{equation}
\begin{aligned}
\chi^-_P(\chi^+_P+\frac{1}{4}\chi_g)+ 2M^2\sum_i\chi^+_{Pi}\chi^-_{Pi}e^{-(M_i-M)^2/s}\Big[\frac{M}{2M_i^3}\\ -\frac{M(M_i-M)}{M_i^2}\Big(\frac{\chi^{'+}_{Pi}}{\chi^+_{Pi}}+\frac{\chi^{'-}_{Pi}}{\chi^-_{Pi}}\Big) -\frac{(M_i-M)^2}{sM_i^2}\Big]\\
+ 2M^2\sum_i\chi_{gi}\chi^-_{Pi}e^{-(M_i-M)^2/s}\Big[\frac{M}{8M_i^3} -\frac{M(M_i-M)}{4M_i^2}\Big(\frac{\chi'_{gi}}{\chi_{gi}}+\frac{\chi^{'-}_{Pi}}{\chi^-_{Pi}}\Big) -\frac{(M_i-M)^2}{4sM_i^2}\Big]\\
= -\frac{7}{2}\frac{M^4}{s^2}(A^u_4-A^d_4)
 +\frac{149}{24}\frac{M^6}{s^3}(A^u_6-A^d_6)  \equiv \Pi^{u+d}_{P4}  .
\end{aligned}
\end{equation}

\begin{figure}[htbp]
\centering
\begin{subfigure}[b]{0.48\textwidth}
\centering
\includegraphics[width=\textwidth]{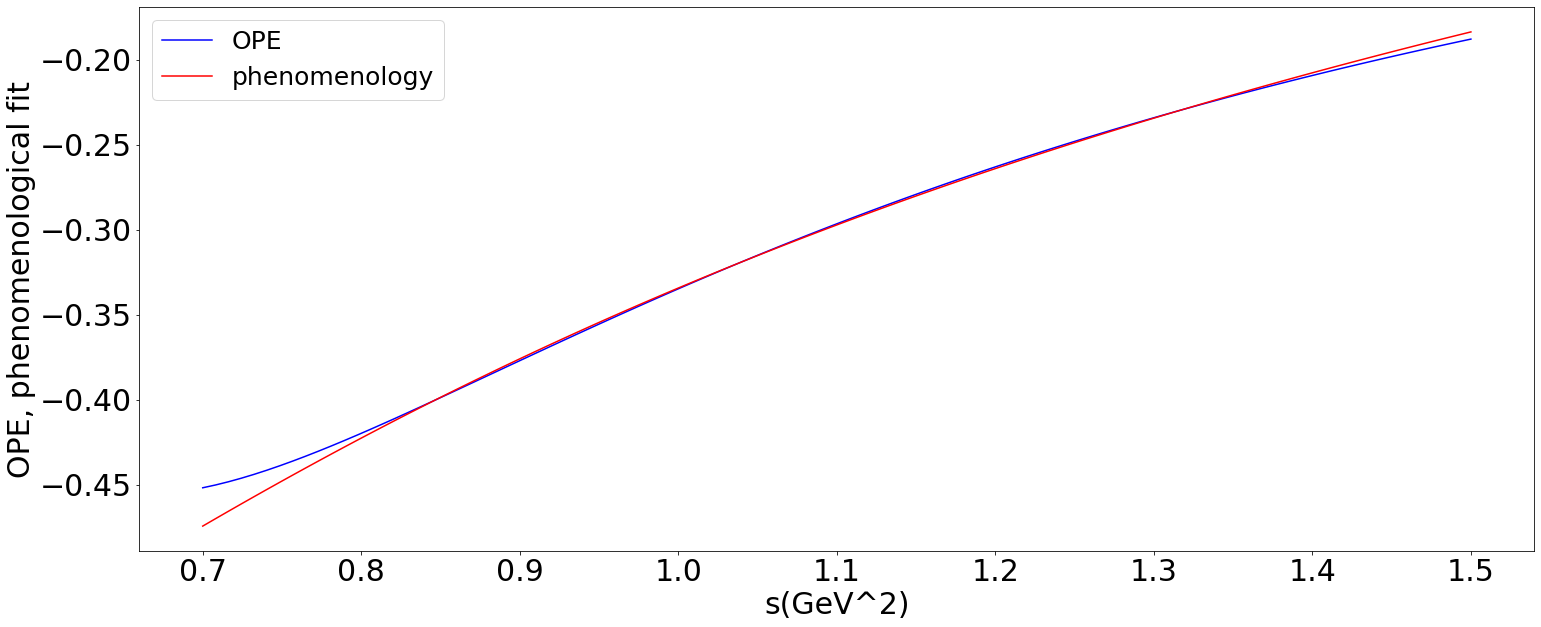}
\caption{{\scriptsize Plots of OPE expression of $\Pi^{u-d}_{41} $ and its fit  as  $0.259-(0.832/s)e^{-0.338/s}$ as functions of s. }}
\label{fig : part1}
\end{subfigure}
\hfill
\begin{subfigure}[b]{0.48\textwidth}
\centering
\includegraphics[width=\textwidth]{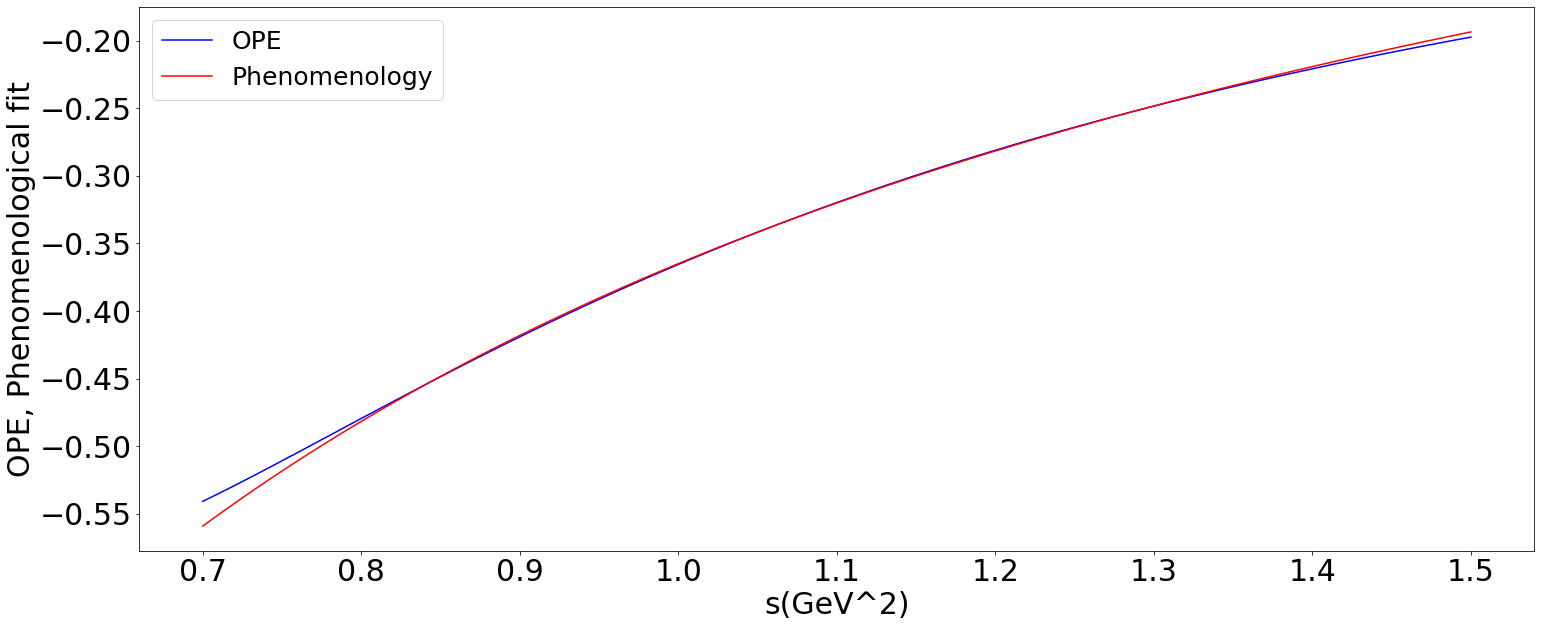}
\caption{{\scriptsize  Plots of OPE expression of $\Pi^{u-d}_{4} $ and its fit  as  $0.210-(0.67/s)e^{-0.153/s}$ as functions of s}}
\label{fig : part2}
\end{subfigure}
\caption{ Two plots, each one of which gives $(\chi_P^-)^2$. } 
\label{fig : main figure}
\end{figure}

From fits of $\Pi^{u-d}_{P41} $ (Fig. 9a) and $\Pi^{u-d}_{P4} $ (Fig. 9b), where fittings are done over the same interval of s, we get $ (\chi^-_P)^2$=(0.259, 0.210) while from fits of $\Pi^{u+d}_{P41} $ (Fig. 10a) and $\Pi^{u+d}_{P4} $ (Fig. 10b), where fittings are done over the same interval of s, we get $ \chi^-_P (\chi^+_P+\chi_g/4)$=(0.119, 0.127) . From these results we conclude : $|\chi^-_P|=0.484\pm 0.026$ and $|\chi^+_P+\chi_g/4|=0.255 \pm 0.022$.

\begin{figure}[htbp]
\centering
\begin{subfigure}[b]{0.48\textwidth}
\centering
\includegraphics[width=\textwidth]{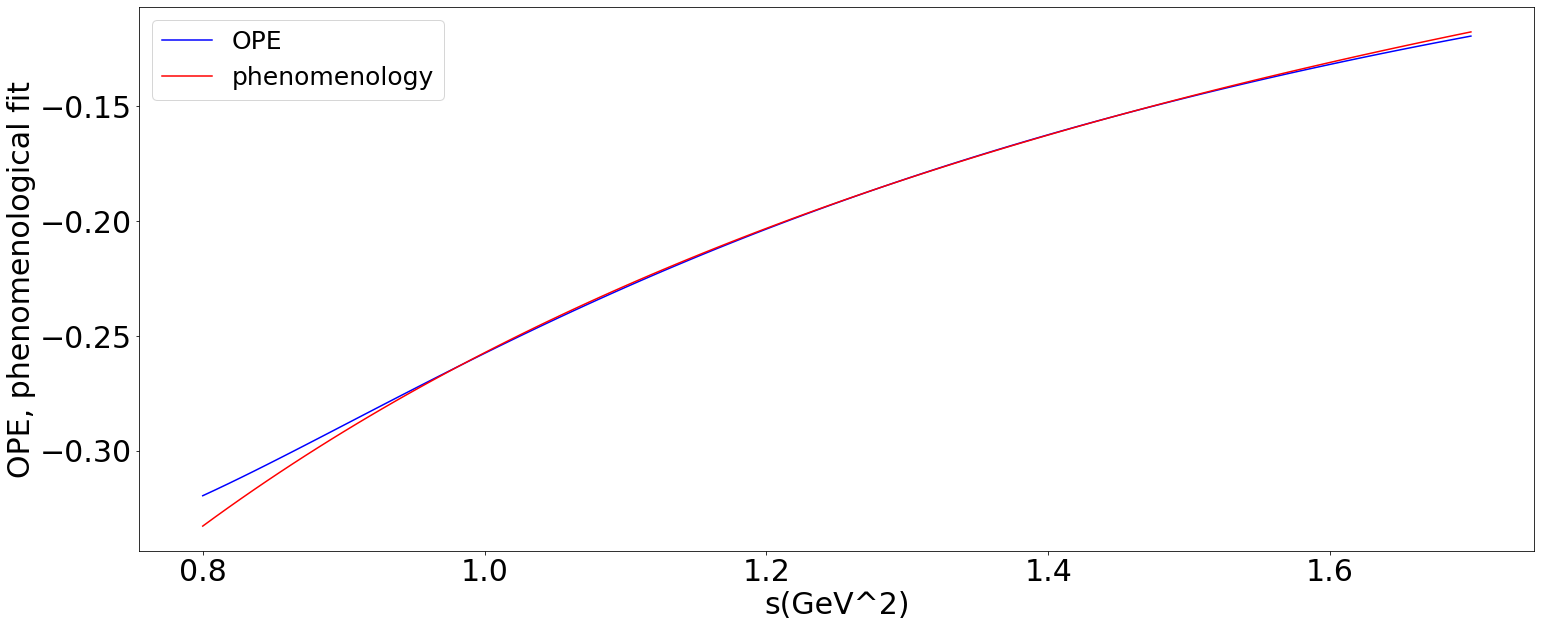}
\caption{{\scriptsize Plots of OPE expression of $\Pi^{u+d}_{41} $ and its fit  as  $0.119-(0.442/s)e^{-0.161/s}$ as functions of s. }}
\label{fig : part1}
\end{subfigure}
\hfill
\begin{subfigure}[b]{0.48\textwidth}
\centering
\includegraphics[width=\textwidth]{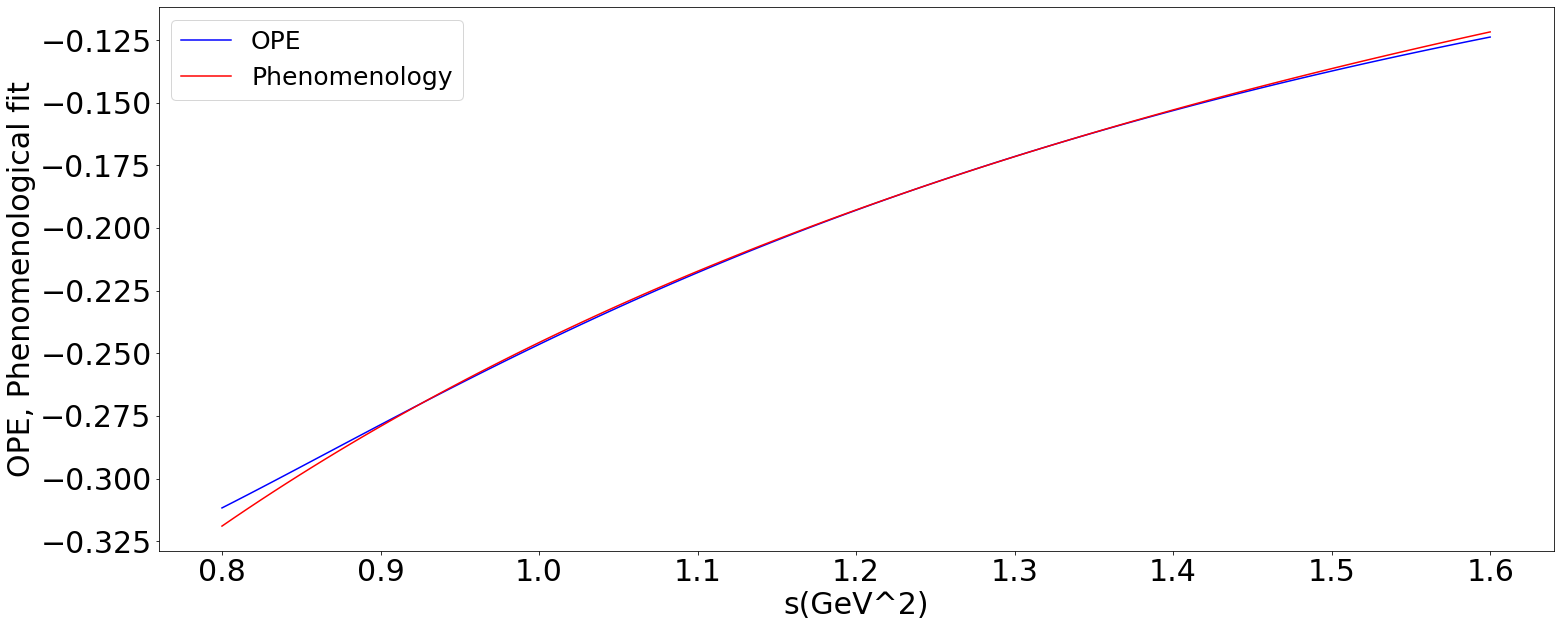}
\caption{{\scriptsize  Plots of OPE expression of $\Pi^{u+d}_{4} $ and its fit  as  $0.127-(0.444/s)e^{-0.175/s}$ as functions of s}}
\label{fig : part2}
\end{subfigure}
\caption{ Two plots, each one of which gives $ \chi^-_P (\chi^+_P+\chi_g/4)$. } 
\label{fig : main figure}
\end{figure}

\begin {equation}
\begin{aligned}
\langle P,S|J^-_{\mu5}|P+q,S' \rangle = \bar{u}(P,S)[g^3_A(q^2)\gamma^{\nu}\gamma_5+h^3_A(q^2)q_{\mu}\gamma_5] u(P+q,S')\\
\langle P,S|J^+_{\mu5}|P+q,S' \rangle = \bar{u}(P,S)[g^{u+d}_A(q^2)\gamma_{\mu}\gamma_5+h^{u+d}_A(q^2)q_{\mu}\gamma_5] u(P+q,S')
\end{aligned}
\end{equation}
On taking divergence of the above equations and taking the limit $q^2 \rightarrow 0$, one gets :\\
\begin {equation}  \chi^-_P=g^3_A/2,     \end{equation} 
\begin {equation}    \chi^+_P+\chi_g/4=g^{u+d}_A/2  \end{equation} 
Similarly, for s-quark alone (see Eq. (44) ) we get
\begin {equation}    \chi_s+\chi_g/8=g^s_A/2  \end{equation} 

 From results obtained, we have $2\chi^-_P=1.018$ to be compared with $g^3_A=1.216$ and $2(\chi^+_P+\chi_g/4)=0.466$ to be compared with $g^{u+d}_A=0.409$. Also, using the result for $\chi_g$ and assuming $\chi_g>0$ we get  $\chi^+_P=0.002\pm0.048$. From Eq. (57) one gets $\chi_s=-0.166$ to be compared with the one obtained from Table III : $\chi_s=-0.213$, where numerically lower value has been taken. \\
\par \hspace{12pt} We also tried to calculate pseudoscalar coupling constants of a nucleon by considering nucleon matrix elements of the correlation functions of two pseudoscalar currents from the beginning without involving axial currents. But in this approach coupling constants come out to be reduced by up to an order of magnitude compared to the results obtained above in this section due to appearance of light quark masses in quadratic form, and naturally they fail internal consistency check.

\section{Summary and conclusion}
In this section we summarize the results,  check their  internal consistency  and compare them with  results in the literature :\\

\begin{table}[h!]
\begin{center}
\caption{{\footnotesize List of results obtained in this work and in some other references. Error bars in numerical values of pseudoscalar couplings arise due to two expressions derived in the text. } }
\label{table:2}
\begin {tabular}{ lc lc lc lc lc lc lc|c| }
\hline \hline
{\footnotesize Ref.$\diagdown$ Parameter} & $|\chi_8|$ & $|\chi_g|$ & $|\chi_P^-|$ & $|\chi_P^++\chi_g/4|$ & $|g_A^3|$ & $|g_A^{u+d}|$ & $|g_A^s|$ \\
\hline
{\footnotesize This work} & {\footnotesize 0.483} & {\footnotesize 1.014} & {\footnotesize 0.484} & {\footnotesize 0.255 } &  {\footnotesize 1.216} &  {\footnotesize0.409} &  {\footnotesize0.054 } \\
  & {\footnotesize $\pm 0.025$} & {\footnotesize $\pm 0.102$} & {\footnotesize $\pm 0.026$} & {\footnotesize $\pm 0.022$} &   &   &   \\
\hline
\cite{Green}(latt.) &   &   &   &   &  {\footnotesize $1.208$} &  {\footnotesize $0.517$}  &  {\footnotesize $0.024$}   \\
  &   &   &   &   &  {\footnotesize $(6)$} &  {\footnotesize $(11)$}  &  {\footnotesize $(21)$}   \\
\hline
\cite{Chiu} &   &   &   &   &    & {\footnotesize $0.35$}   &   \\
\hline
\cite{JPS1} &   &   &   &   &    &  {\footnotesize $0.46$}   &   \\
\hline
\cite{Tomalak} &   &   &   &   & {\footnotesize1.265(26)}  &   &   \\
\hline
\cite{Bali} {\footnotesize (calculated)} &   &   & {\footnotesize $0.674$}   &   &    &    &   \\
\hline
\cite{Deur} &   &   &   &   &    &    &  {\footnotesize $ 0.05(5)$}  \\
\hline
\cite{Kaplan} &   &   &   &   &    &    &  {\footnotesize $ 0.15(8)$}  \\
\hline
\cite{Barone}(latt.) &   &   &   &   &    &    &  {\footnotesize $ 0.0325(78)$} \\
\hline \hline
\end{tabular}
\end{center}
\end{table}

\begin{table}[h!]
\begin{center}
\caption{{\footnotesize List of some parameters calculated from those  given in Table II. We assume $g^3_A>0$, $g^{u+d}_A>0$. The result for the last column has been obtained ausuming  $\chi_s<0$ as shown in the text and errors from different sources have been added in the same direction.} }
\label{table:3}
\begin {tabular}{ lc lc lc lc lc lc lc|c| }
\hline \hline
{\footnotesize Ref.$\diagdown$ Parameter} & $g_A^u$ & $g_A^d$ & $\chi_P^+$ & $\chi_s $ & $ 2\chi_P^- $ & $ {\footnotesize 2(\chi_P^++\frac{\chi_g}{4})} $ & $ {\footnotesize 2(\chi_P^++\chi_s)+\frac{3}{4}\chi_g} $ \\
\hline
This work & {\footnotesize$ 0.813$} & {\footnotesize $-0.404$} & {\footnotesize $0.002 $} & {\footnotesize $-0.252$ } &  {\footnotesize0.968} &  {\footnotesize0.464} &  {\footnotesize $0.261 \pm 0.251$}  \\
   &   &   & {\footnotesize $\pm 0.048$} & {\footnotesize $\pm 0.039 $ } & {\footnotesize $\pm 0.052$ }  & {\footnotesize $\pm 0.056$ }  &   \\
\hline
\cite{Blumlein} &  {\footnotesize 0.866}  &   {\footnotesize -0.404}   &   &   &    &   &   \\
\hline
\cite{Sato} &  {\footnotesize 0.83(1)}  &  {\footnotesize -0.44(1)}   &   &   &    &   &   \\
\hline
\cite{Green} &  {\footnotesize 0.863}  &  {\footnotesize -0.345}   &   &   &    &   &   \\
    &  {\footnotesize (7)(14)}  &  {\footnotesize (6)(9)}   &   &   &    &   &   \\
 \hline \hline
\end{tabular}
\end{center}
\end{table}
From $|\chi_8|$, $|\chi_g|$ and $|\chi_P^++\chi_g/4|$, we can get $\chi_P^+$ and $\chi_s $ assuming $\chi_8>0$ and $\chi_g>0$ : $\chi_P^+=0.002 \pm 0.048$, $\chi_s=-0.252 \pm 0.039$. \\
On taking one nucleon matrix elements of Eq. (7)  divergence equations we get :
\begin{equation} g_A^8=2 \chi_8 =2(\chi_P^+-2\chi_s), g_A^0  \equiv g_A^{u+d+s}=2(\chi_P^++\chi_s)+\frac{3}{4}\chi_g \end{equation}

From Eq.(33) and assuming $|\chi_8|$=0.458, the lower value, $g_A^8$ should be 0.91; however, $g_A^8=0.58\pm 0.03$ \cite{Close}  assuming SU(3) symmetry and $|g_A^8|\simeq 0.59$ \cite{JPS1} obtained in the same approach as the current work sec. 3. Clearly the  relation between axial coupling and the corresponding pseudoscalar coupling works better for the isovector current compared to the octet current. The reason for this lies in our taking limit $q^2 \rightarrow 0$ to get the relation. The physically measured values of axial couplings apply at physical points. While the isovector axial current couples to $\pi^0$, the octet axial current couples largely to $\eta$ and to some extent to $\eta'$ due to mixing. Due to larger masses of the latter, the limit $q^2 \rightarrow 0$ gives rise to more discrepancy for the octet case.   Also due to its dominant role in $\chi_8$, $\chi_s < 0$ if $g_A^8>0$. Adding all the errors from different sources in the same direction
\begin{equation} \chi_P^++\chi_s=-0.250\pm 0.087 \end{equation}
\begin{equation} 2(\chi_P^++\chi_s)+\frac{3}{4}\chi_g= 0.261 \pm 0.251 \end{equation}
From Eqs. (58) and (60), $g_A^0\equiv g_A^{u+d+s}=  0.261 \pm 0.251$. This may be  compared to the COMPASS result : $g_A^0= 0.32\pm0.02\pm0.04\pm0.05$ \cite{Adolph} as well as lattice results at $\mu=2 GeV$ : 0.494(11)(15) \cite{Green}, 0.355(43) \cite{Barone}.\\
\par \hspace{12pt} From Ref. \cite{Bali} : \\
\begin {equation} g_A=1.25, g_P=280(27)     \end{equation} 
For comparision, our result for $\chi^-_P=(\bar{m}/(2M))g_P= 0.674$ (taking only the lower value in Eq. (61)) as compared to 0.509 obtained here.
\par \hspace{12pt} Errors  in pseudoscalar coupling constants listed in Table II are $\lesssim$ 10\%. Other than that, I estimate errors in various couplings due to variation in the range of Borel mass parameter to be around 15\% and errors due to variation in QCD parameters to be around 10\% making them a total of 25\% error.  

\par \hspace{12pt} In summary, in this paper we have calculated pseudoscalar and axial coupling constants of a nucleon in a approach where one-nucleon matrix elements of correlation functions of pseudoscalar or axial currents are calculated in terms of nucleon matrix elements of quark, gluon and quark-gluon composite operators and moments of parton distribution function. Furthermore, in this work we have taken into account contributions of excited states and continuum in the phenomenological expression of the correlation function through an exponential parameterization, a step not taken in the previous works in this approach \cite{Nishikawa, Nishikawa1, JPS1}.  For the pseudoscalar coupling constants we derived two different expressions which were giving approximately same numerical results. One interesting observation is that one of the expressions for pseudoscalar coupling  constants of the nucleon contained only moments of parton distribution function. That makes them independent of other QCD parameters as well as quark masses, and hence will remain unchanged in the chiral limit. $\chi^+_P $, $\chi_s $  and $\chi^+_P +\chi_s$, in association with $\chi_g$, have been related to the corresponding nucleon axial coupling constant $g^{u+d}_A$, $g^s_A$ and $g^0_A$, and these are largely found to be consistent. This makes us to believe that the numerical value of $\chi_g$ is approximately correct. We also checked the internal consistency of pseudoscalar coupling constant of the nucleon $\chi^-_P$  and the corresponding axial coupling constants of the nucleon $g^3_A$ and found that they were largely consistent.\\
\par \hspace{12pt} We believe our results on nucleon matrix elements of pseudoscalar currents, and in particular of  axial anomaly, largely ignored in the current literature, will find application in hadron physics.
\section{ACKNOWLEDGEMENT}
I thank Prof. J. Pasupathy for suggesting the problem, for having numerous discussions and for his keen interest to see this work completed.

\bibliographystyle{unsrt}
\bibliography{comp_ax31_ar}

\appendix
\section{ Appendix : From axial-axial current correlator to axial anomaly-pseudoscalar current correlator}
In this Appendix we derive Eq. (9) which relates one-nucleon matrix element of an axial anomaly-octet pseudoscalar current correlator to one-nucleon matrix element of axial-axial current correlators. 
With reference to \cite{Cheng} we can write
\begin{equation} 
\begin{aligned}
\int d^4xe^{iqx}[\langle N(p)|T\{\partial^{\mu}J_{\mu 5}^0(x), \partial^{\nu}J_{\nu 5}^8(0)\}|N(p)\rangle \\
=q^{\mu}q^{\nu}\int d^4xe^{iqx}\langle N(p)|  T\{J_{\mu 5}^0(x), J_{\nu 5}^8(0)\}|N(p)\rangle  \\ 
+\int d^4xe^{iqx}\delta (x_0)\{iq^{\mu}\langle N(p)| [J_{05}^8(0),J_{\mu 5}^0(x)] |N(p)\rangle  \\ 
-\langle N(p)| [J^0_{05}(x), \partial^{\mu}J_{\mu 5}^8(0)] |N(p)\rangle\}
\end{aligned}
\end{equation}
It is to be noted that this applies when operators are sandwiched between one-nucleon (hadron) states and it is not an operator relation. It is easy to see that
\begin{equation} \delta (x_0) [\bar{u}(0)\gamma_0 \gamma_5u(0), \bar{u}(x)\gamma_{\mu} \gamma_5u(x)]=0 \end{equation}
which implies
\begin{equation}\delta (x_0) [J_{05}^8(0),J_{\mu 5}^0(x)]=0, \delta (x_0) [J_{05}^8(0),J_{\mu 5}^8(x)]=0 \end{equation}
and
\begin{equation}\delta (x_0) [\bar{u}(x)\gamma_0 \gamma_5u(x), \bar{u}(0) \gamma_5u(0)]=-2\delta^4 (x)\bar{u}(0)u(0)\end{equation}
implies
\begin{equation} \delta (x_0) [J^0_{05}(x), \partial^{\mu}J_{\mu 5}^8(0)]=-\frac{4i}{3\surd{2}}\delta^4 (x)[m_u\bar{u}(0)u(0)+m_d\bar{d}(0)d(0)-2m_s\bar{s}(0)s(0)] \end{equation}
Similarly,
\begin{equation} 
\begin{aligned}
\int d^4xe^{iqx}[\langle N(p)|T\{\partial^{\mu}J_{\mu 5}^8(x) \partial^{\nu}J_{\nu 5}^8(0)\}|N(p)\rangle \\
=q^{\mu}q^{\nu}\int d^4xe^{iqx}\langle N(p)|  T\{J_{\mu 5}^8(x)J_{\nu 5}^8(0)\}|N(p)\rangle  \\ 
+\int d^4xe^{iqx}\{iq^{\mu}\delta (x_0)\langle N(p)| [J_{05}^8(0),J_{\mu 5}^8(x)] |N(p)\rangle  \\ 
-\delta (x_0)\langle N(p)| [J^8_{05}(x), \partial^{\mu}J_{\mu 5}^8(0)] |N(p)\rangle\},
\end{aligned}
\end{equation}

Also, use the following for the next step
\begin{equation} \delta (x_0) [J^8_{05}(x), \partial^{\mu}J_{\mu 5}^8(0)]=-\frac{2i}{3}\delta^4 (x)[m_u\bar{u}(0)u(0)+m_d\bar{d}(0)d(0)+4m_s\bar{s}(0)s(0)] \end{equation}
Next, consider
\begin{equation} 
\begin{aligned}
2\surd{2}\partial^{\mu}J_{\mu 5}^0(x) \partial^{\nu}J_{\nu 5}^8(0)-\partial^{\mu}J_{\mu 5}^8(x) \partial^{\nu}J_{\nu 5}^8(0) \\
=Q(x)P_8(0)+2(m_u\bar{u}(x)i\gamma_5u(x)+m_d\bar{d}(x)i\gamma_5d(x))\times \\ (m_u\bar{u}(0)i\gamma_5u(0)+m_d\bar{d}(0)i\gamma_5d(0)) \\ -8m_s^2\bar{s}(x)i\gamma_5s(x) \bar{s}(0)i\gamma_5s(0)+4m_s \bar{s}(x)i\gamma_5s(x)\times \\
(m_u\bar{u}(0)i\gamma_5u(0)+m_d\bar{d}(0)i\gamma_5d(0)) \\
-4(m_u\bar{u}(x)i\gamma_5u(x)+m_d\bar{d}(x)i\gamma_5d(x))m_s  \bar{s}(0)i\gamma_5s(0) \\
\cong Q(x)P_8(0) -8m_s^2\bar{s}(x)i\gamma_5s(x) \bar{s}(0)i\gamma_5s(0)
\end{aligned}
\end{equation} 
The coefficient of of $m_s m_{u,d}$vanishes upto four-quark operators, and hence has been neglected. Hence,
\begin{equation} 
\begin{aligned}
\int d^4xe^{iqx} \langle N(p)| T\{Q(x),P_8(0)\} -8m_s^2T\{\bar{s}(x)i\gamma_5s(x), \bar{s}(0)i\gamma_5s(0)\} |N(p) \rangle \\
\cong \int d^4xe^{iqx} \langle N(p)|T \{2\surd{2}\partial^{\mu}J_{\mu 5}^0(x), \partial^{\nu}J_{\nu 5}^8(0)\}-T\{\partial^{\mu}J_{\mu 5}^8(x), \partial^{\nu}J_{\nu 5}^8(0)\} |N(p) \rangle \\
=2\surd{2} q^{\mu}q^{\nu}\int d^4xe^{iqx}\langle N(p)|  T\{J_{\mu 5}^0(x),J_{\nu 5}^8(0)\}|N(p)\rangle  \\ 
+\frac{8i}{3}\int d^4xe^{iqx}\delta^4(x)\langle N(p)|(m_u\bar{u}(0)u(0)+m_d\bar{d}(0)d(0)-2m_s\bar{s}(0)s(0))|N(p)\rangle  \\
- [q^{\mu}q^{\nu}\int d^4xe^{iqx}\langle N(p)|  T\{J_{\mu 5}^8(x),J_{\nu 5}^8(0)\}|N(p)\rangle  \\ +\frac{2i}{3} \int d^4xe^{iqx}\delta^4(x)\langle N(p)|(m_u\bar{u}(0)u(0)+m_d\bar{d}(0)d(0)+4m_s\bar{s}(0)s(0))|N(p)\rangle]
\end{aligned}
\end{equation}

\begin{equation} 
\begin{aligned}
i\int d^4xe^{iqx} \langle N(p)| T\{Q(x),P_8(0) \}|N(p)\rangle \\
\cong i8m_s^2\int d^4xe^{iqx} \langle N(p)| T\{\bar{s}(x)i\gamma_5s(x), \bar{s}(0)i\gamma_5s(0)\} |N(p) \rangle + \\
q^{\mu}q^{\nu} i\int d^4xe^{iqx} \langle N(p)|2\surd{2}T\{J_{\mu 5}^0(x),J_{\nu 5}^8(0)\}-T\{J_{\mu 5}^8(x),J_{\nu 5}^8(0)\} |N(p)\rangle \\
-2 \langle N(p)|(m_u\bar{u}(0)u(0)+m_d\bar{d}(0)d(0)-4m_s\bar{s}(0)s(0))|N(p)\rangle
\end{aligned}
\end{equation}
Though not explicitly written, subtraction of $\langle ... \rangle_0 \langle p,s|p,s \rangle $ is understood in each of the terms under integral. 

\section{Appendix : Axial anomaly-octet pseudoscalar current correlator in coordinate space}
In this Appendix we write the expression for the axial anomaly-octet pseudoscalar current correlator in coordinate space assuming that ultimately it will be Fourier transformed with respect to variable $q^{\mu}$ and sandwiched between one-nucleon state. This is because the axial-axial current correlator is already multiplied with $q^{\mu} q^{\nu}$ (see Eqs. (8) and (9) and the following equation).
\begin{equation}
\begin{aligned}
\Pi^{(08)}(x,q) \equiv   T\{Q(x),P_8(0)\}=2\sqrt{2}q^{\mu}q^{\nu}   T\{J^{(0)}_{\mu5}(x), J^{(8)}_{\nu5}(0)\} \\-T\{J^{(8)}_{\mu5}(x), J^{(8)}_{\nu5}(0)  \} +8m_s^2 T\{\bar{s}(x)i\gamma_5s(x),\bar{s}(0)i\gamma_5s(0)\} \\
-2(m_u\bar{u}u+m_d\bar{d}d-4m_s\bar{s}s)\\
=\frac{1}{2}\Big[\frac{-1}{48\pi^2}\big(\frac{q^2}{x^2}+\frac{2(q.x)^2}{x^4}\big)\frac{\alpha_s}{\pi}G^2+\frac{1}{4\pi^2}\big(\frac{3q^2}{x^2}-\frac{2(q.x)^2}{x^4}\big) \times \\  (m_u\bar{u}u+m_d\bar{d}d-4m_s\bar{s}s) +\frac{i}{\pi^2}\big(q^2\frac{x^{\mu}x^{\nu}}{x^4}  -2\frac{q.xq^{\mu}x^{\nu}}{x^4}\big)\times \\
(\bar{u}\gamma_{\mu}D_{\nu}u+\bar{d}\gamma_{\mu}D_{\nu} d-4\bar{s}\gamma_{\mu}D_{\nu} s)-\frac{i}{6\pi^2x^4}x^{\alpha}x^{\beta}x^{\lambda}x^{\sigma} (2q_{\sigma}q_{\mu}-q^2g_{\sigma \mu}) \times \\ (\bar{u}\gamma^{\mu}D_{\alpha}D_{\beta}D_{\lambda}u+\bar{d}\gamma^{\mu}D_{\alpha}D_{\beta}D_{\lambda}d-4\bar{s}\gamma^{\mu}D_{\alpha}D_{\beta}D_{\lambda} s) \\
-\frac{g^2}{144\pi^2}\big(2\frac{(q.x)^2}{x^2}-q^2\big)(\bar{u}\gamma^{\lambda}t^a u+\bar{d}\gamma^{\lambda}t^a d-4\bar{s}\gamma^{\lambda}t^a s)\sum_q\bar{\psi}_q\gamma_{\lambda}t^a\psi_q \\
-\frac{g^2}{48\pi^2}\big[\frac{x^{\mu}x^{\nu}}{x^4}+\frac{1}{2}\frac{g^{\mu \nu}}{x^2}\big] (\bar{u}\gamma_{\mu}\lambda^a u+\bar{d}\gamma_{\mu}\lambda^a d-4\bar{s}\gamma_{\mu}\lambda^a s)\sum_q\bar{\psi}_q\gamma_{\nu}\lambda^a\psi_q \\    -\frac{\Gamma(d/2-2)m_s}{3456\pi^2(-x^2)^{(d/2-2)}}[2(q.x)^2-q^2x^2]\bar{s}sg^2G^2 +\frac{\Gamma(d/2-2)\Gamma(d/2-1)}{648(16\pi^2)^2(-x^2)^{(d-4)}}\\ \times [2(q.x)^2-q^2x^2](g^2G^2)^2  -8\frac{\alpha_s}{\pi}m_s^2 ln(-x^2)(\bar{s}t^a\gamma_5s)(\bar{s}t^a\gamma_5s) \\   -\frac{i}{120\pi^2x^4}x^{\alpha}x^{\beta}x^{\lambda}x^{\sigma}x^{\rho}x^{\nu}(2q_{\mu}q_{\nu} -q^2g_{\mu \nu})(\bar{u}\gamma^{\mu}D_{\alpha}D_{\beta}D_{\lambda}D_{\sigma}D_{\rho}u+\\ \bar{d}\gamma^{\mu}D_{\alpha}D_{\beta}D_{\lambda}D_{\sigma}D_{\rho}d-4\bar{s}\gamma^{\mu}D_{\alpha}D_{\beta}D_{\lambda}D_{\sigma}D_{\rho} s)   \Big]\\
+8m_s^2\Big[\frac{-g^2G^2}{128\pi^4x^2}-\frac{i}{\pi^2}\frac{x^{\mu}x^{\nu}}{x^4}\bar{s}\gamma_{\mu}D_{\nu}s -\frac{i}{6\pi^2x^4}x^{\alpha}x^{\beta}x^{\lambda}x^{\sigma}\bar{s}\gamma_{\sigma}D_{\alpha}D_{\beta}D_{\lambda}s \\   -\frac{g^2}{6\pi^2}\big[\frac{1}{16}\frac{x^{\mu}x^{\nu}}{x^2}-\frac{x^{\mu}x^{\nu}}{x^4 q^2}\big] (\bar{s}\gamma_{\mu}\lambda^a s)\sum_q\bar{\psi}_q\gamma_{\nu}\lambda^a\psi_q \\     +\frac{\alpha_s}{8\pi}\big[ln(-x^2)(\bar{s}\lambda^a\sigma^{\mu \nu}s)(\bar{s}\lambda^a\sigma_{\mu \nu}s)+2\frac{x_{\mu}x^{\nu}}{x^2}\big(\frac{2}{q^2x^2}-1\big)\times \\(\bar{s}\lambda^a\sigma^{\alpha \mu}s)(\bar{s}\lambda^a\sigma_{\alpha \nu}s)\big]\Big] -2(m_u\bar{u}u+m_d\bar{d}d-4m_s\bar{s}s) 
\end{aligned}
\end{equation}
where, as per our scheme, perturbative terms are not included and d=dimension of space-time. 

\end{document}